\documentclass[onecolumn]{aa}
\usepackage[caption=false]{subfig}
\usepackage[colorlinks,linkcolor=blue,anchorcolor=blue,citecolor=blue]{hyperref}
\usepackage[varg]{txfonts}
\DeclareMathAlphabet{\mathcal}{OMS}{cmsy}{m}{n}
\DeclareSymbolFont{largesymbols}{OMX}{cmex}{m}{n}

\newcommand{\obs}{\theta_{\rm{obs}}}
\newcommand{\jet}{\theta_{\rm{j}}}
\newcommand{\jc}{\theta_{\rm{c}}}
\newcommand{\np}{\nu^{\prime}}
\newcommand{\nc}{\nu_{\rm{c}}^{\prime}}
\newcommand{\nmin}{\nu_{\rm{min}}^{\prime}}
\newcommand{\nmax}{\nu_{\rm{max}}^{\prime}}
\newcommand{\ys}{\frac{1}{3}}
\newcommand{\ye}{\frac{1}{2}}
\newcommand{\pe}{\frac{p}{2}}
\newcommand{\ype}{\frac{1-p}{2}}
\newcommand{\fmax}{F_{\np,\rm{max}}}
\newcommand{\Ek}{E_{\rm{k,iso}}}
\newcommand{\ts}{t_{\rm{s}}}
\newcommand{\ncc}{\nu_{\rm{c}}}
\newcommand{\nmm}{\nu_{\rm{min}}}

\begin{document}
	
\title{Possible origin of AT2021any: a failed GRB from a structured jet  }

\author{Fan Xu\inst{1},
	Yong-Feng Huang\inst{1,2,3}\thanks{e-mail: {\tt hyf@nju.edu.cn}}, and
	Jin-Jun Geng\inst{4}
}

\institute{
	\inst{1}{School of Astronomy and Space Science, Nanjing University, Nanjing 210023, People's Republic of China}\\
    \inst{2}{Xinjiang Astronomical Observatory, Chinese Academy of Sciences, Urumqi 830011, People's Republic of China}\\
	\inst{3}{Key Laboratory of Modern Astronomy and Astrophysics (Nanjing University), Ministry of Education, People's Republic of China}\\
	\inst{4}{Purple Mountain Observatory, Chinese Academy of Sciences, Nanjing 210023, People's Republic of China}\\
}

\date{}
	
\abstract {Searching for afterglows not associated with any gamma-ray
bursts (GRBs) is a longstanding goal of transient surveys. These surveys
provide the very chance of discovering the so-called orphan afterglows.
Recently, a promising orphan afterglow candidate, AT2021any, was found
by the Zwicky Transient Facility. Here we perform multi-wavelength
fitting of AT2021any with two different outflow models, namely the
top-hat jet model and the structured
Gaussian jet model. Although the two models can both fit the observed
light curve well, it is found that the structured Gaussian jet model
presents a better result, and thus is preferred by observations.
In the framework of the Gaussian jet model, the best-fit Lorentz
factor is about $68$, which indicates that AT2021any should be a
failed GRB. The half-opening angle of the jet and the viewing angle
are found to be 0.104 and 0.02, respectively, which means that the
jet is essentially observed on-axis. The trigger time of the GRB
is inferred to be about $1000$ s before the first detection of the
orphan afterglow. An upper limit of $21.5\%$ is derived for the
radiative efficiency, which is typical in GRBs. }


\titlerunning{Orphan afterglow AT2021any}
\authorrunning{Xu et al.}
\maketitle

\section{Introduction} \label{sec:intro}

Gamma-ray bursts (GRBs) are energetic explosions occurring at
cosmological distances. There are mainly two different phases in
GRBs, the prompt emission and the follow-up afterglow phase \citep{Piran..2004,Meszaros..2006,Kumar..2015}. The
prompt emission usually peaks at sub-MeV \citep{Fishman..1995,Gruber..2014}, while the afterglow can
be observed in a much broader wavelength ranging from soft X-rays
to radio waves \citep{Meszaros..1997,Sari..1998,Zhang..2006,Kann..2010,Geng..2016,Alexander..2017,Connor..2023}. Generally speaking,
the afterglow can last for a period ranging from days to years,
whereas the prompt emission typically lasts for less than a few
minutes. Despite the short duration of the prompt emission, the
total energy released in $\gamma$-rays is enormous. For long-duration
GRBs, the typical isotropic energy of the prompt emission is $10^{53}$
erg \citep{Xu..2023}. Recently, the discovery of GRB 221009A has once
again refreshed our perception of the energetics of
GRBs \citep{Ren..2023,An..2023,Connor..2023,Sato..2023,Yang..2023}.
This event is found to have an isotropic energy of $\sim 10^{55}$ erg,
which places it as the most energetic GRB in history \citep{An..2023,Yang..2023}.

For a typical GRB, both the prompt emission and the afterglow phases
can be observed. However, there exist special cases of GRBs where the
prompt emission is missing and only the afterglow phase could be observed.
These afterglows are the so-called orphan
afterglows \citep{Rhoads..1997,Huang..2002,Nakar..2002,Zou..2007,Gao..2022}.
Theoretically, there are two different approaches to producing an orphan
afterglow. First, if the initial Lorentz factor of the outflow is
significantly less than $\sim 100$, the prompt emission will be too faint
to be observed. Such explosions are also known as ``failed GRBs'' or ``dirty
fireballs'' \citep{Paczynski..1998,Dermer..1999,Huang..2002,Xu..2012}.
Second, if the outflow is highly collimated (not isotropic), it is hard
to detect the prompt emission when the explosion is viewed off-axis,
i.e., the viewing angle is larger than the half-opening angle of the
jet \citep{Rhoads..1997,Nakar..2002}. In this case, the majority of
$\gamma$-ray radiation from the GRB cannot be observed due to the
relativistic beaming effect. However, the subsequent afterglow
emission will be less beamed and can be visible to the observer
in the afterglow stage \citep{Rhoads..1997,Huang..2002}. These
transients are called off-axis orphans.

It is also interesting to note a class of X-ray transients
named X-ray flashes. Their temporal and spectral features of
X-ray emissions are very similar to those of normal GRBs. The main
difference is that X-ray flashes usually possess a significantly lower
spectral peak energy than GRBs in the prompt emission phase.
X-ray flashes can also be explained by the dirty fireball
model \citep{Huang..2002} or the off-axis model \citep{Yamazaki..2002},
which means they may be closely connected with orphan
afterglows \citep{Urata..2015}.

Assuming that all the orphan afterglows come from off-axis GRBs,
we can estimate the half-opening angle of GRBs through the ratio
between the orphan afterglow rate and GRB
rate \citep{Rhoads..1997,Paczynski..2000}. However, as pointed
out by \citet{Huang..2002}, there should be many orphan afterglows
coming from failed GRBs since the baryon loading issue widely exists
in popular progenitor models of GRBs \citep{Piran..2004}. Therefore
the estimation above may not be very straightforward.

On the other hand, searching for orphan afterglows is not an easy
task \citep{Grindlay..1999,Greiner..2000,Levinson..2002,Rau..2006,Gal..2006,
Khabibullin..2012,Ho..2018,Huang..2020,Ho..2022,Leung..2023}. So far, only six orphan afterglow candidates have been reported.
One of them was found at radio wavelength by the Very Large
Array (VLA): FIRST J$141918.9+394036$ \citep{Law..2018,Mooley..2022}.
Its radio evolution and the lack of associated GRB suggest that it
might be an off-axis orphan afterglow \citep{Mooley..2022}. The other
five transients are all found in optical bands. They are
PTF11agg \citep{Cenko..2013,Wang..2013,Wu..2014} discovered by the
Palomar Transient Factory \citep{Law..2009}, and AT2019pim
\citep{Kool..2019}, AT2020blt \citep{Ho..2020,Andreoni..2021,Sarin..2022},
AT2021any \citep{Ho..2022,Gupta..2022}, and
AT2021lfa \citep{Ho..2022,Lipunov..2022} discovered by the Zwicky
Transient Facility (ZTF; \citet{Bellm..2019}).

With the currently available observational data, it is still a challenge
to derive the physical parameters and reveal the nature of an orphan
afterglow. A major problem is the lack of the trigger time of the
unseen GRB potentially associated with the afterglow. Usually,
the trigger time is estimated by fitting the observed
light curve with a particular model \citep{Ho..2020,Gupta..2022,Ho..2022}.
Recently, \cite{Sarin..2022} used the last non-detection information from
an upper limit in the $r$ band to constrain the trigger time. However, the
non-detection may result from bad seeing or other interferences, thus
cannot provide decisive information on the trigger time.

In this study, we present an in-depth study on AT2021any,
an orphan afterglow candidate found by ZTF. The trigger time,
together with other parameters, will be derived by fitting the
observed light curve. An efficient code is developed for this
purpose. Synchrotron emission and synchrotron self-Comptonization
(SSC) are considered in our modeling. The effect of synchrotron
self-absorption is also included.

Our paper is organized as follows. First, the multi-wavelength
observational data of AT2021any are collected and presented in
Section \ref{sec:obs}. The physical models used to fit the data
are then briefly described in Section \ref{sec:mod}. In
Section \ref{sec:fit}, we present the fitting results of AT2021any
with different approaches and compare the goodness of fitting.
Finally, conclusions and discussion are presented in Section \ref{sec:dis}.

\section{Observational data of AT2021any/ZTF21aayokph} \label{sec:obs}

AT2021any/ZTF21aayokph was discovered on $06$:$59$:$45.6$ UTC $2021$
January $16$ by ZTF \citep{Ho..2021a}. It had an $r$ band magnitude of
$r = 17.92 \pm 0.06$ mag when it was first detected. The most recent
non-detection was only $20.3$ minutes before the first detection \citep{Ho..2022},
which gives a limiting magnitude of $r > 20.28$ mag. No associated
GRB was recorded during the period between the last non-detection and
the first detection \citep{Ho..2022,Gupta..2022}. The object faded rapidly
in the $r$ band, with a fading rate of $14$ mag day $^{-1}$ during the
first $3.3$ hours. The extinction-corrected color index was found to
be $g - r = (0.25 \pm 0.19)$ mag \citep{Ho..2022}. These observations
place AT2021any as a promising orphan afterglow candidate. Using
spectroscopic observations, the redshift of AT2021any was later
determined as $z=2.5131 \pm 0.0016$ \citep{Postigo..2021,Ho..2022}.
AT2021any was subsequently followed by a variety of optical facilities
(see \cite{Ho..2022} and \cite{Gupta..2022} for more information).
The multi-wavelength optical photometry data are collected and listed
in Table \ref{tab:1}. Note that the AB magnitudes in Table \ref{tab:1}
are not corrected for the extinction of the Milky Way.

The object was followed in X-rays by Swift-XRT \citep{Ho..2021b}. The
observations were performed in three different epochs, with a total
exposure time of $8.2$ ks. X-ray emission was detected only in the
first epoch. The unabsorbed flux density is estimated
as $3.30 \times 10^{-13}$ erg cm$^{-2}$ s$^{-1}$, with a neutral
hydrogen column density
of $N_{\rm{H}} = 8.12 \times 10^{20}$ cm$^{-2}$ \citep{Willingale..2013}
and an assumed photon index of $\Gamma_{\rm{p}} = 2$  (see Table \ref{tab:2}).

The transient was observed in radio by VLA \citep{Perley..2021,Ho..2022}.
Eight epochs of observations were performed from $4.90$ days to $75.77$
days after the discovery of AT2021any. We have collected the radio data
obtained by \cite{Ho..2022} and listed them in Table \ref{tab:3}.

\section{Dynamics and emission mechanisms of GRBs} \label{sec:mod}

In this section, we briefly describe the dynamic evolution and
radiation process of relativistic outflows that produce GRBs.
The dynamics of the outflow can be depicted by the following
equations \citep{Huang..2000a,Huang..2006,Geng..2013,Xu..2022}
\begin{eqnarray}
	&& \frac{dR}{dt} = \beta c \Gamma (\Gamma + \sqrt{\Gamma ^{2} - 1}), \\
	&& \frac{dm}{dR} = 2\pi R^{2} (1 - \cos\jet) n m_{\rm{p}}, \label{eq:1}\\
	&& \frac{d\Gamma}{dm} = -\frac{\Gamma^{2} - 1}{M_{\rm{ej}}
		+ \epsilon_{\rm{r}} m + 2(1 - \epsilon_{\rm{r}}) \Gamma m } .
\end{eqnarray}
Note that the lateral expansion of the outflow is neglected in our
calculation. Here, $R$ is the shock radius in the GRB rest frame, $c$ is the
speed of light, and $\Gamma$ is the Lorentz factor of the outflow
with $\beta = \sqrt{\Gamma ^{2}-1}/ \Gamma$. $t$ is the observer's
time. $m$ is the swept-up mass of the interstellar medium (ISM),
$\jet$ is the half-opening angle of the outflow, and $m_{\rm{p}}$
is the proton's mass. $n$ is the number density of the surrounding
ISM. For a homogeneous ISM, we take $n$ as a constant. As for a
wind ISM, we have $n = Ar^{-2}$, where $A$ is a coefficient
depending on the mass loss rate and the speed of the
wind \citep{Chevalier..1999,Dai..2001,Wu..2003,Ren..2023}.
$M_{\rm{ej}}$ is the initial mass of the ejecta
and $\epsilon_{\rm{r}}$ is the radiative efficiency.
In this study, note that when we calculate
the dynamical evolution of a structured jet, our method is to
divide the whole jet into many mini-jets. For a mini-jet located
at angle $\theta$, Equation \ref{eq:1} should then be modified
as $\frac{dm}{dR} = \sin \theta d\theta d\phi R^{2} n m_{\rm{p}}$ to
calculate the swept-up mass, where $d\theta$ is the angular length
of the mini-jet and $d\phi$ is its width.

Synchrotron radiation from shock-accelerated electrons is
involved in GRB afterglows \citep{Sari..1998}. We use the
superscript prime ($\prime$) to denote the quantities in the
shock comoving frame, while those without prime are in the
observer frame. In the fast-cooling regime, the flux $F_{\np}$
at frequency $\np$ is
\begin{equation} \label{eq:2}
	F_{\np} = \fmax \left\{ \begin{array}{ll}
			(\frac{\np}{\nc})^{\ys}, & \np < \nc, \\
			(\frac{\np}{\nc})^{\ye}, & \nc < \np < \nmin, \\
			(\frac{\np}{\nmin})^{-\pe}(\frac{\nc}{\nmin})^{\ye}, & \nmin < \np < \nmax,
							\end{array} \right.
\end{equation}
where $\nc$, $\nmin$, and $\nmax$ are characteristic frequencies
corresponding to the cooling Lorentz factors $\gamma_{c}^{\prime}$,
the minimum Lorentz factor $\gamma_{\rm{min}}^{\prime}$, and maximum
Lorentz factor $\gamma_{\rm{max}}^{\prime}$, respectively. $p$ is
the electron spectral index and $\fmax$ is the peak flux
density \citep{Sari..1998}.

In the slow-cooling regime, when $\nmin < \np < \nmax$, we have
\begin{equation} \label{eq:3}
	F_{\np} = \fmax \left\{ \begin{array}{ll}
		(\frac{\np}{\nmin})^{\ys}, & \np < \nmin, \\
		(\frac{\np}{\nmin})^{\ype}, & \nmin < \np < \nc, \\
		(\frac{\np}{\nc})^{-\pe}(\frac{\nc}{\nmin})^{\ype}, & \nc < \np < \nmax.
	\end{array} \right.
\end{equation}
In the case of $\nmax < \nc $, we have
\begin{equation} \label{eq:4}
	F_{\np} = \fmax \left\{ \begin{array}{ll}
		(\frac{\np}{\nmin})^{\ys}, & \np < \nmin, \\
		(\frac{\np}{\nmin})^{\ype}, & \nmin < \np.
	\end{array} \right.
\end{equation}

We use the Compton parameter $Y$ to denote the ratio of the inverse
Compton scattering luminosity with respect to synchrotron luminosity.
It can be calculated as
\begin{equation} \label{eq:5}
	Y = \frac{-1+\sqrt{1+4\epsilon_{\rm{r,e}} \epsilon_{\rm{e}} \epsilon_{B} }}{2}.
\end{equation}
Here $\epsilon_{\rm{r,e}}$ is the fraction of the electron energy
that was radiated, while $\epsilon_{\rm{e}}$ is the fraction of
thermal energy carried by electrons and $\epsilon_{B}$ is the
ratio of magnetic field energy to the total energy \citep{Sari..2001,Wei..2006}.

The synchrotron self-absorption effect is also considered in our
calculations. As a result, the observed flux should be corrected
by multiplying a factor
of $f(\tau_{\nu^{\prime}})=(1-e^{-\tau_{\nu^{\prime}}}) / \tau_{\nu^{\prime}}$,
where $\tau_{\nu^{\prime}}$ is the optical depth. The self-absorption
coefficients and the optical depth are calculated by
following \cite{Wu..2003} and \cite{Geng..2016}.

The observed flux density at frequency $\nu$ can be calculated as
\begin{equation} \label{eq:6}
	F_{\nu}(\Theta) = (1+z)\mathcal{D}^{3} f(\tau_{\nu^{\prime}}) F_{\np},
\end{equation}
where $\Theta$ stands for the angle between the velocity of emitting
material and the line of sight. $\mathcal{D}=1/[\Gamma(1-\beta \cos\Theta)]$
is the Doppler factor and $\nu^{\prime}=(1+z) \nu / \mathcal{D}$.
To calculate the observed flux $F_{\nu}(t)$ at a given time $t$, we
need to integrate the emission power over the equal arrival time
surface (EATS), which is determined by
\begin{equation} \label{eq:7}
	t=(1+z)\int\frac{1-\beta \cos \Theta}{\beta c}dR.
\end{equation}

\section{Multi-wavelength fitting of the afterglow} \label{sec:fit}

In this section, we present multi-wavelength fitting of AT2021any by
considering two different models. First, a simple top-hat jet model is
used, which has a constant energy per solid angle and a uniform Lorentz
factor within the jet. Secondly, we consider a structured jet model
with a Gaussian profile \citep{Kumar..2003,Troja..2018,Geng..2019,Lamb..2021}.
In this case, the distribution of the kinetic energy is taken
as $E(\theta)= \Ek \exp(-\theta^{2}/\jc^{2})$ at angle $\theta$,
and the profile of the Lorentz factor is assumed to take the form
of $\Gamma(\theta)=(\Gamma_0 - 1)\exp(-\theta^{2}/2\jc^{2}) + 1$.
Here $\Ek$ is the isotropic equivalent energy on the jet axis ($\theta = 0$)
and $\Gamma_0$ is the corresponding Lorentz factor. $\jc$ stands for the
half-opening angle of the jet core.
For convenience, we assume that the jet is cut off at an angle of
$\jet$, i.e., $E(\theta)=0$ and $\Gamma(\theta)=1$ for $\theta>\jet$.
In other words, $\jet$ effectively denotes the edge of the structured
jet.

The observed broadband X-ray flux data are converted to the flux at
the frequency of $\nu=1 \times 10^{18}$ Hz. To do so, a photon index
of $\Gamma_{\rm{p}} = 2$ is applied \citep{Ho..2022}. In the optical
bands, we consider a Galactic extinction of $E(B - V) = 0.0575$
mag \citep{Schlafly..2011} to convert the observed magnitudes to
flux densities. The radio afterglow data in four different bands are
also used in our fitting, i.e., S(3GHz), C(6GHz), X(10GHz), and Ku(15GHz)
bands.

We use the Markov Chain Monte Carlo (MCMC) algorithm to get the best-fit
results for the multi-wavelength afterglow of AT2021any. The
parameters derived for different models are shown in Table \ref{tab:4}.
The shift time of the light curve $\ts$ is defined as the time interval
between the GRB trigger and the first observation. This parameter is
introduced to find out the most probable trigger time of the GRB
associated with AT2021any. A constant ISM density is considered here
for simplicity.

\subsection{The top-hat jet model}

\begin{figure*}
	\centering
	\subfloat{
		\includegraphics[width=\textwidth]{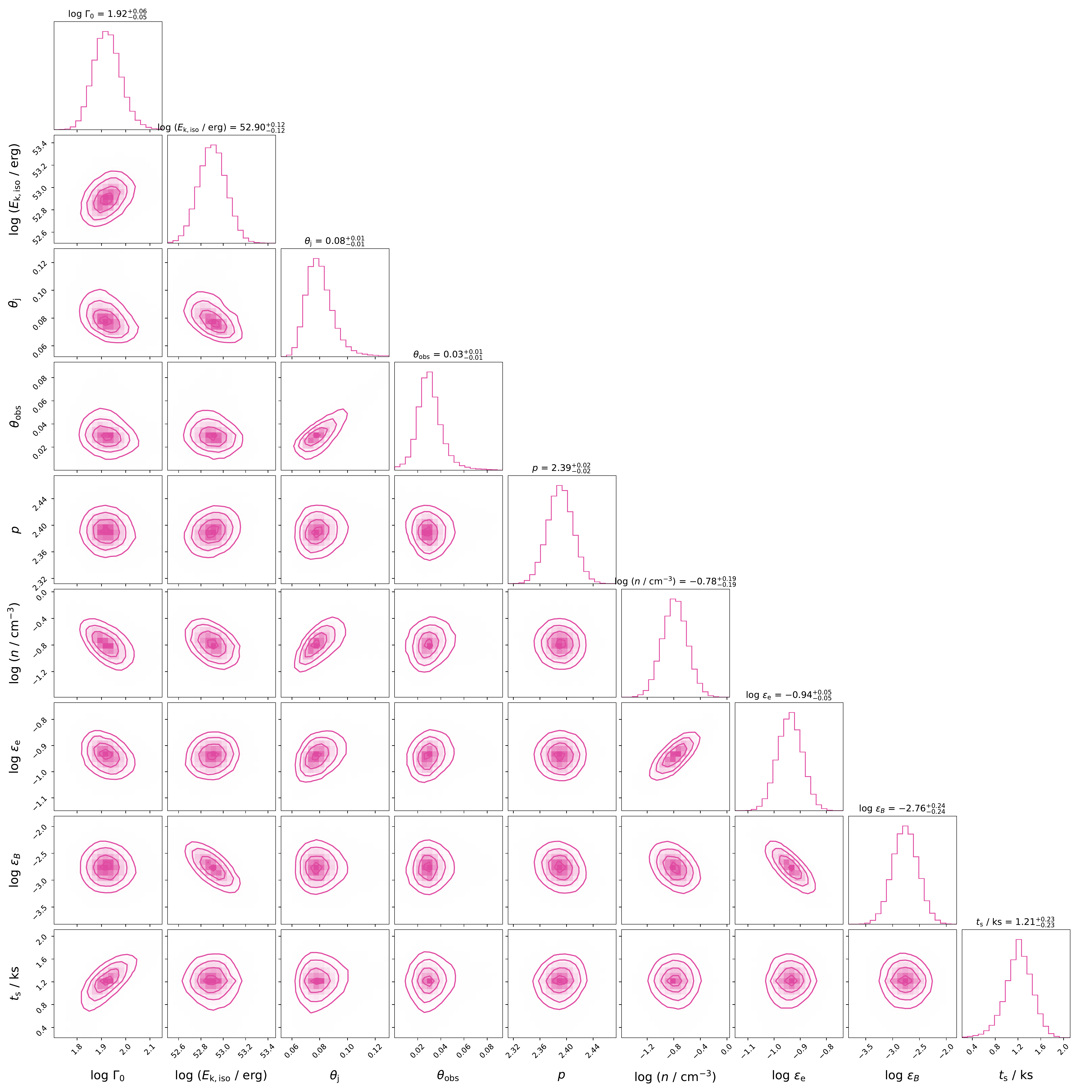}
	}
	
	\caption{Parameters derived for AT2021any by using the top-hat jet
             model ($1\sigma-3\sigma$ confidence levels are shown).
             The best-fit results are marked with $1\sigma$
             uncertainties above the panel of their posterior
             distribution.}   \label{fig:1}
\end{figure*}

\begin{figure*}
	\centering
	\subfloat{
		\includegraphics[width=\textwidth]{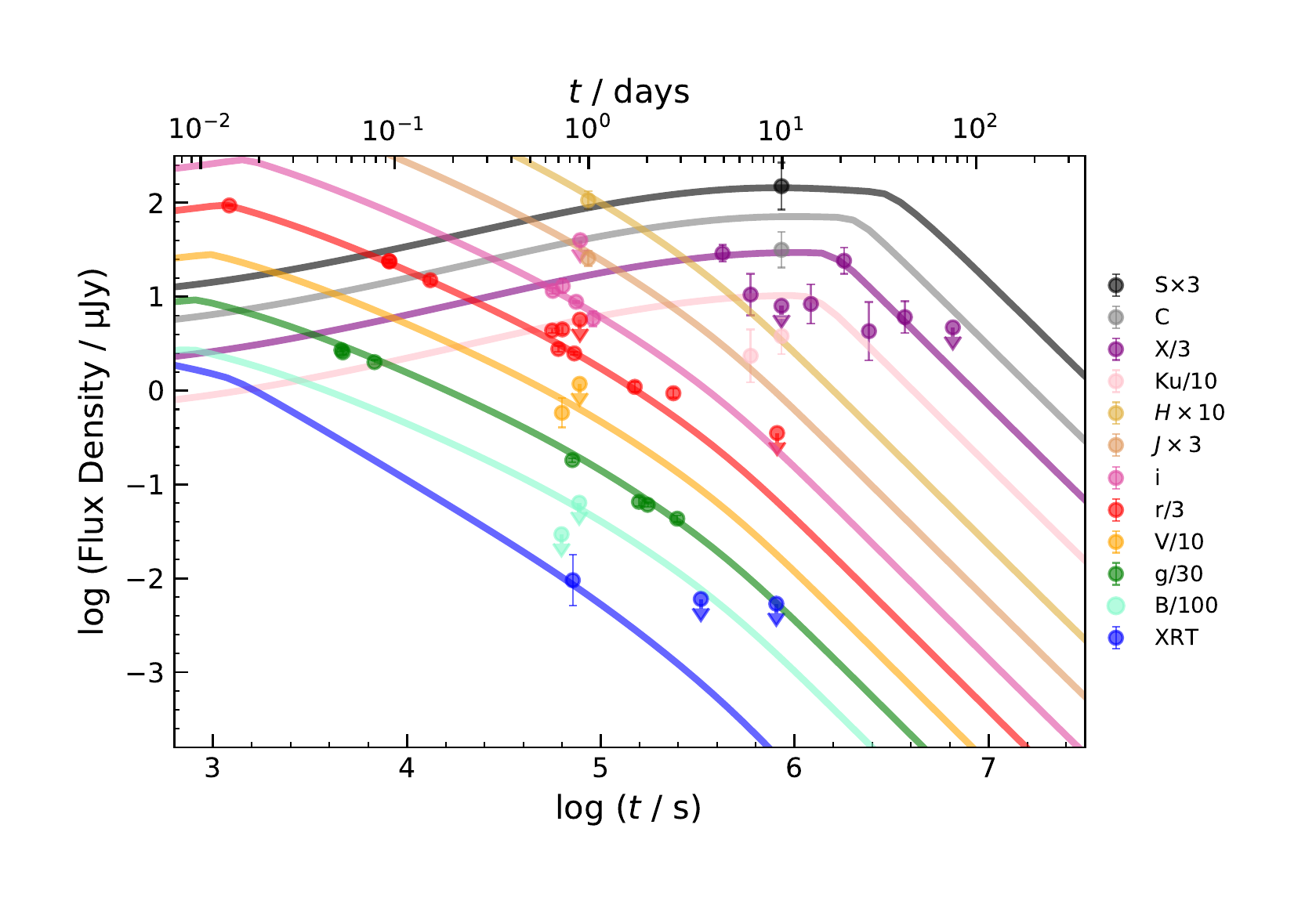}
	}
	
	\caption{Observed multi-wavelength afterglow of AT2021any and
             the best-fit result by using the top-hat jet model
             (solid curves). The dashed line represents the broken power law fitting result of the $r$ band data. } \label{fig:2}
\end{figure*}

We begin our fitting with the top-hat jet model. The numerical results are
presented in Table \ref{tab:4} and the corresponding corner plot is shown in
Figure \ref{fig:1}. The best-fit value for the initial Lorentz factor of
the jet ($\Gamma_{0}$) is $\sim 83$, which favors a failed GRB origin.
Note that the Lorentz factor is generally sensitive to the early afterglow,
especially its onset. In the case of AT2021any, although the onset of the
afterglow was not clearly detected, the most recent non-detection was luckily
only 20.3 minutes before the first detection. It gives a firm
constraint on the onset of the afterglow. This is the reason that the
Lorentz factor can be effectively inferred from the observations.
The multi-wavelength observational data points
and the best-fit light curves are plotted in Figure \ref{fig:2}. We see
that the X-ray and optical data are generally well-fitted. However,
the radio data seem to somewhat deviate from the
theoretical light curves. Note that the effect of interstellar scintillation
is not included in our calculations. Small-scale inhomogeneities in the
ISM can cause scintillation by changing the phase of radio waves. The
line of sight to a distant source also shifts as the Earth moves,
leading to radio flux fluctuations. The scattering effect will
be significant when the radio frequency is smaller than the transition
frequency. In fact, \cite{Ho..2022} pointed out that
interstellar scintillation may have a significant contribution to the
radio afterglow of AT2021any. They calculated the transition
frequency at the direction of AT2021any and got a result of $15$ GHz
by using the NE2001 model \citep{Cordes..2002}. Therefore, the radio
light curves we considered here could be largely affected by
interstellar scintillation.

In Figure \ref{fig:2}, we see that the optical light cures
have an obvious break at about $t_{\rm{b}} \sim 0.4$ days. Before this
time, the temporal index is about $\alpha_{\rm{opt},1}= -0.75 \pm 0.1$, while
it becomes $\alpha_{\rm{opt},2}= -1.33 \pm 0.23$ after the break time. Note that
$\alpha_{\rm{opt},2}$ is satisfactorily consistent with the result expected for
the fireball model \citep{Sari..1998}, i.e. $-(3p-2)/4 = -1.3$ for $p \sim 2.4$
from our best fit. It indicates that the optical band has crossed the cooling
frequency at around $0.4$ days. The theoretical temporal index before the
optical band crosses the cooling frequency should be $-3(p-1)/4 = -1.05$. It is
smaller than the observed value of $\alpha_{\rm{opt},1}= -0.75$ as mentioned
above. The difference may be due to the EATS effect, which is more significant
at the early stages of the afterglow.

In Figure \ref{fig:2}, the half-opening angle of the jet is found to
be $0.08^{+0.01}_{-0.01}$, while the viewing angle is $0.03^{+0.01}_{-0.01}$.
It indicates that we are essentially observing the jet on the axis. An achromatic
jet break could be seen in the optical light curves and the temporal decay
index is around $-2$ after the jet break. The break time is about $\sim 2$ days,
which is also roughly consistent with the theoretically jet break time
of $t_{\rm{j}} \sim 1.2\frac{(1+z)}{2} E_{\rm{k,iso}, 53}^{1/3} n^{-1/3}
\theta_{\rm{j}, -1}^{8/3} = 1.96$ days \citep{Sari..1999}.

Note that the decay index of our theoretical X-ray light curve is
around $1.3$ between $0.01$ and $1$ day. It is obviously steeper
than the optical light curves in the same period. This is easy to understand.
The frequencies of X-ray photons are much higher than that of optical photons.
As a result, it is in the fast cooling regime (i.e., the frequency is higher
than the characteristic cooling frequency). Then, the theoretical decay index
should be $\sim -(3p-2)/4 = 1.3$ \citep{Sari..1998}, which is well consistent
with our numerical result. SSC might have some effects on the X-ray afterglow.
Here we present some further discussions on this issue. The effect of SSC can
be assessed by the Compton parameter $Y$, which is defined as the ratio of
the inverse Compton scattering luminosity with respect to the synchrotron
luminosity. As shown in Eq.~(\ref{eq:5}), $Y$ is sensitively dependent
on $\epsilon_{\rm{r,e}}$, i.e., the fraction of the radiated electron
energy. According to \citet{Sari..2001}, $\epsilon_{\rm{r,e}}$ takes the form
of $\epsilon_{\rm{r,e}} = (\frac{\gamma_{\rm{min}}}{\gamma_{c}})^{p-2}$
considering that the bulk of electrons are in the slow-cooling regime at
the afterglow phase. It can be further expressed as $\epsilon_{\rm{r,e}}
= (1.27 \times 10^{-8} \frac{p-2}{p-1} \epsilon_{\rm{e}}
\epsilon_{B} \Gamma^{4} t)^{p-2}$ \citep{Huang..2000b}.
Taking $\Gamma \propto t^{-3/8}$ for the adiabatic expansion
case \citep{Sari..1998} and combining the best-fit parameters for the
top-hat jet discussed here, we finally have $\epsilon_{\rm{r,e}} \sim 0.007 t^{-0.2}$.
Consequently, we get $Y = \frac{-1+\sqrt{1+4.76 t^{-0.2}}}{2}$
from Eq.~(\ref{eq:5}). We see that the value of $Y$ will decrease
from $Y = 0.35$ at $t = 100$ s to $Y = 0.16$ at $t = 10000$ s.
Therefore SSC will generally have a negligible effect on the afterglow
of AT2021any.

The radio light curve peaks at about $20$ days. The pre-peak temporal
index of the radio light curve is about $1/2$. It indicates that the radio
emission will reach the peak flux when the observed radio frequency crosses
the characteristic frequency of $\nmm$ \citep{Zhang..2018}. The post-peak
radio light curve is dominated by the jet break effect. Here we further
address the effect of synchrotron self-absorption, which is mainly determined
by the synchrotron self-absorption frequency ($\nu_{\rm{a}}$).
Following \cite{Gao..2013}, we can derive the self-absorption frequency
as $\nu_{\rm{a}} = 1.03 \times 10^{9}$ Hz in the case of $\nu_{\rm{a}} < \nmm < \ncc$.
On the other hand, it will be $\nu_{\rm{a}} = 8.3 \times 10^{12} (\frac{t}{s})^{-0.72}$ Hz
in the case of $\nmm < \nu_{\rm{a}} < \ncc$. In both cases, we see
that $\nu_{\rm{a}}$ is in the radio ranges. Thus the synchrotron
self-absorption will mainly affect the radio afterglow light curves
and will have a negligible effect on the optical and X-ray light curves.


\subsection{The structured jet model}

\begin{figure*}
	\centering
	\subfloat{
		\includegraphics[width=\textwidth]{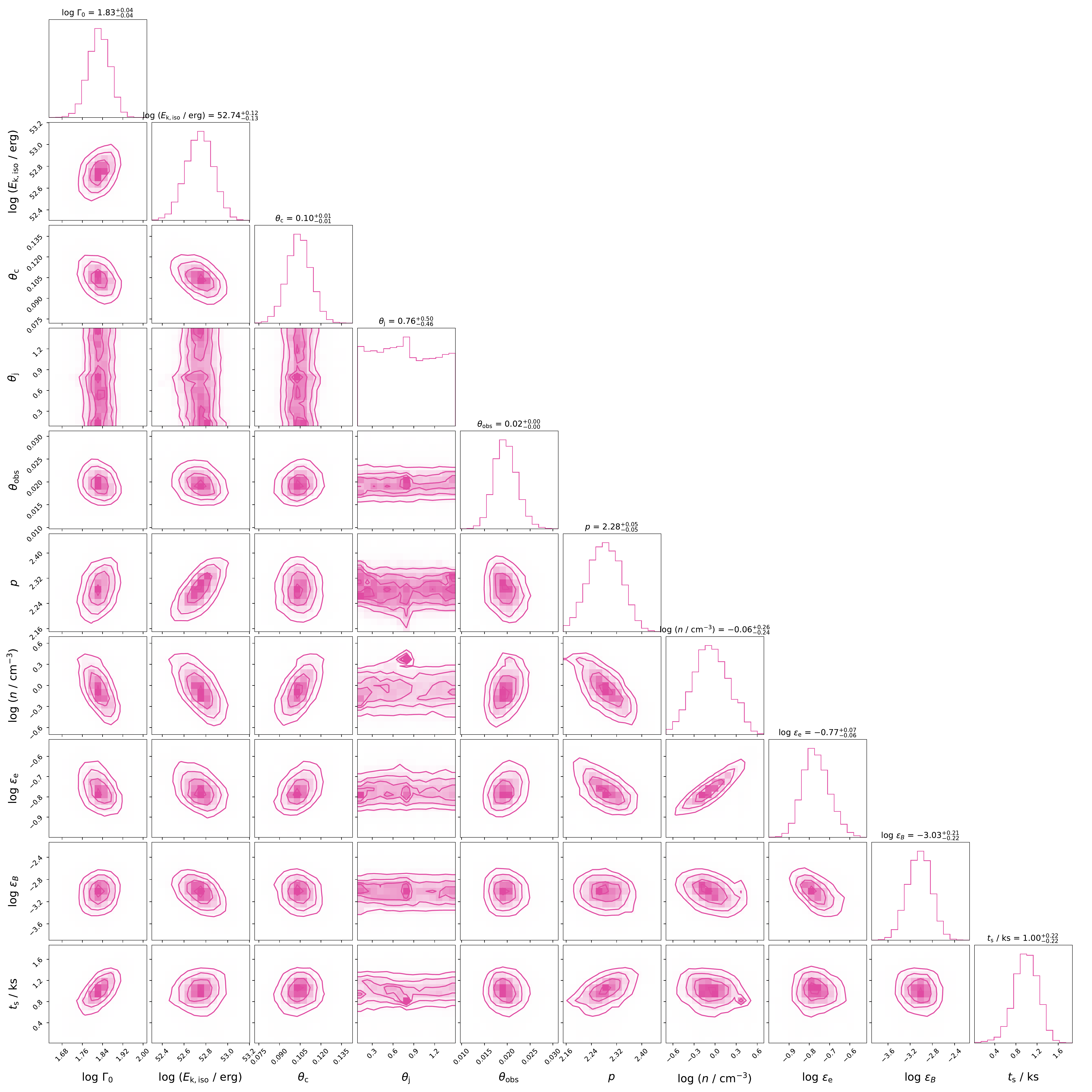}
	}
	
	\caption{Parameters derived for AT2021any by using the structured jet
             model ($1\sigma-3\sigma$ confidence levels are shown).
             The best-fit results are marked with $1\sigma$
             uncertainties above the panel of their posterior
             distribution.} \label{fig:3}
\end{figure*}

We have also tried to fit the observational data with a structured jet.
The best-fit results are presented in Table \ref{tab:4}, and the corresponding
corner plot is shown in Figure \ref{fig:3}. We see that the best-fit $\Gamma_{0}$
value is $\sim 68$, which is still a typical Lorentz factor for a failed GRB.
The geometry parameters are derived as $\jc=0.10\pm0.01$,
$\jet=0.76^{+0.50}_{-0.46}$, and $\obs=0.02^{+0.003}_{-0.002}$. An on-axis
viewing angle is still favored here. Figure \ref{fig:3} shows that the error
bar of $\jet$ is relatively large. This is due to the fact that the radiation
from materials outside the jet core contributes little to the observed emissions,
which means that the observed flux is insensitive to $\jet$. In fact, at the early
stage, we could only see a small fraction of the jet due to the beaming effect.
As the jet decelerates, the Lorentz factor is decreasing and we could see a
larger area of the jet. However, the Lorentz factor of the materials outside
the jet core will be too small to produce significant emissions at later stages,
leading to a steep decay in the afterglow light curve. We
derive $\Ek \sim 5.50 \times 10^{52}$ erg, $p \sim 2.3$, $\epsilon_{\rm{e}} \sim 0.17$,
and $\epsilon_{B} \sim 0.001$ for the structured jet model. These parameter values
are similar to those of the top-hat jet model. As for the ambient density $n$,
the structured jet model requires a relatively larger value of $n \sim 0.87$ cm$^{-3}$
as compared to $n \sim 0.16$ cm$^{-3}$ for the top-hat jet model. It indicates
that a cleaner circum-burst environment is needed for the top-hat jet model.
In the framework of the structured jet model, the best-fit shift time is
about $\ts \sim 1000$ s, which is $\sim 200$ s smaller than that obtained
for the top-hat jet model. 

\begin{figure*}
	\centering
	\subfloat{
		\includegraphics[width=\textwidth]{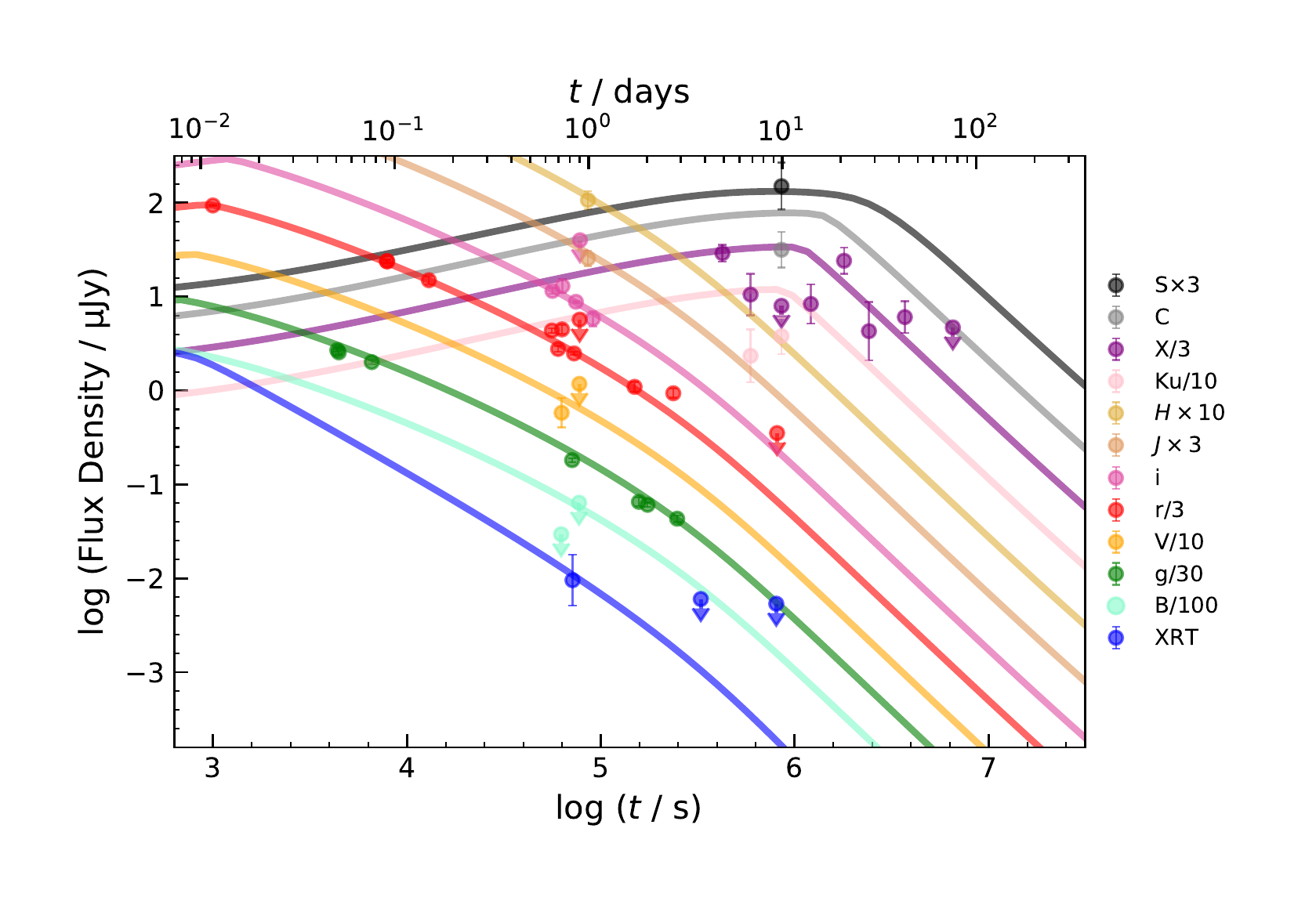}
	}
	
	\caption{Observed multi-wavelength afterglow of AT2021any and
             the best-fit result by using the structured jet model
             (solid curves). The dashed line represents the broken power law fitting result of the $r$ band data. } \label{fig:4}
\end{figure*}

The observational data points are compared with the best-fit light
curves of the structured jet model in Figure \ref{fig:4}. Similar to Figure \ref{fig:2},
we see that the observed radio data points show significant fluctuations and
thus could not be satisfactorily fit by the theoretical light curves (especially
in the C band). Again, it may be due to the interstellar scintillation effect.
The theoretical $r$ band optical light curve still possesses a shallow decay,
with a timing index of $-0.69\pm0.01$ before $t_{\rm{b}} \sim 0.35$ days.
It is slightly larger than the analytical value of $-3(p-1)/4 = -0.975$ and
may be due to the EATS effect. After $t_{\rm{b}} \sim 0.35$ days, the timing
index is $-1.33$, which is roughly consistent with the analytical result
of $\sim -(3p-2)/4 = -1.225$. The jet break time is about $2$ days for
the structured jet model.

\subsection{Comparing the goodness of fitting}

We have assessed the goodness of fitting for different models. Two tests
are performed for this purpose. First, we use the reduce-$\chi^{2}$,
which is calculated as
\begin{equation} \label{eq:8}
	\chi^{2}/\rm{d.o.f.}=\sum_i \frac{ (\log f_{\rm{th},i}
    - \log f_{\rm{obs},i})^{2} }{\sigma_i^2} \times \frac{1}{ \rm{ d.o.f.} },
\end{equation}
where the degree of freedom (d.o.f.) is defined as the difference between
the number of observational data points and the model parameters.
$f_{\rm{th},i}$ and $f_{\rm{obs},i}$ are the theoretical flux density and
the observed flux density at the time of $t_{i}$, while $\sigma_i$
represents the error bar for each data point. The reduced-$\chi^{2}$
for each model is presented in Table \ref{tab:4}. We find that the
structured jet model has a relatively lower reduced-$\chi^{2}$, suggesting
that it is the preferred model.

The second test is conducted by performing the Bayesian Information Criterion
(BIC) method \citep{Schwarz..1978}. BIC is defined as
\begin{equation} \label{eq:9}
	\rm{BIC} = -2\ln \mathcal{L} (P) + k \ln (N),
\end{equation}
where $N$ is the number of observational data points and $k$ is the number
of model parameters. Here $P$ stands for a set of the model parameters
and $\mathcal{L}$ is the maximized value of the likelihood function. The
likelihood function takes the form of \citep{Xu..2021}
\begin{equation} \label{eq:10}
	\mathcal{L}(P)=\prod_{i}
	\frac{1}{\sqrt{2\pi}\sigma_i} \exp\left[-\frac{1}{2} \frac{ (\log f_{\rm{th},i} - \log f_{\rm{obs},i})^{2} }{\sigma_i^2}
	 \right].
\end{equation}
According to the BIC test, the model which provides the minimum BIC score
should be the preferred model. Usually, the BIC score is compared
through $\Delta$BIC values, i.e., the difference between the best model and
other models. We list the $\Delta$BIC score for each model in
Table \ref{tab:4}. Again we see that the structured jet scenario is better than the top-hat jet scenario.

\begin{figure*}
	\centering
	\subfloat{
		\includegraphics[width=\textwidth]{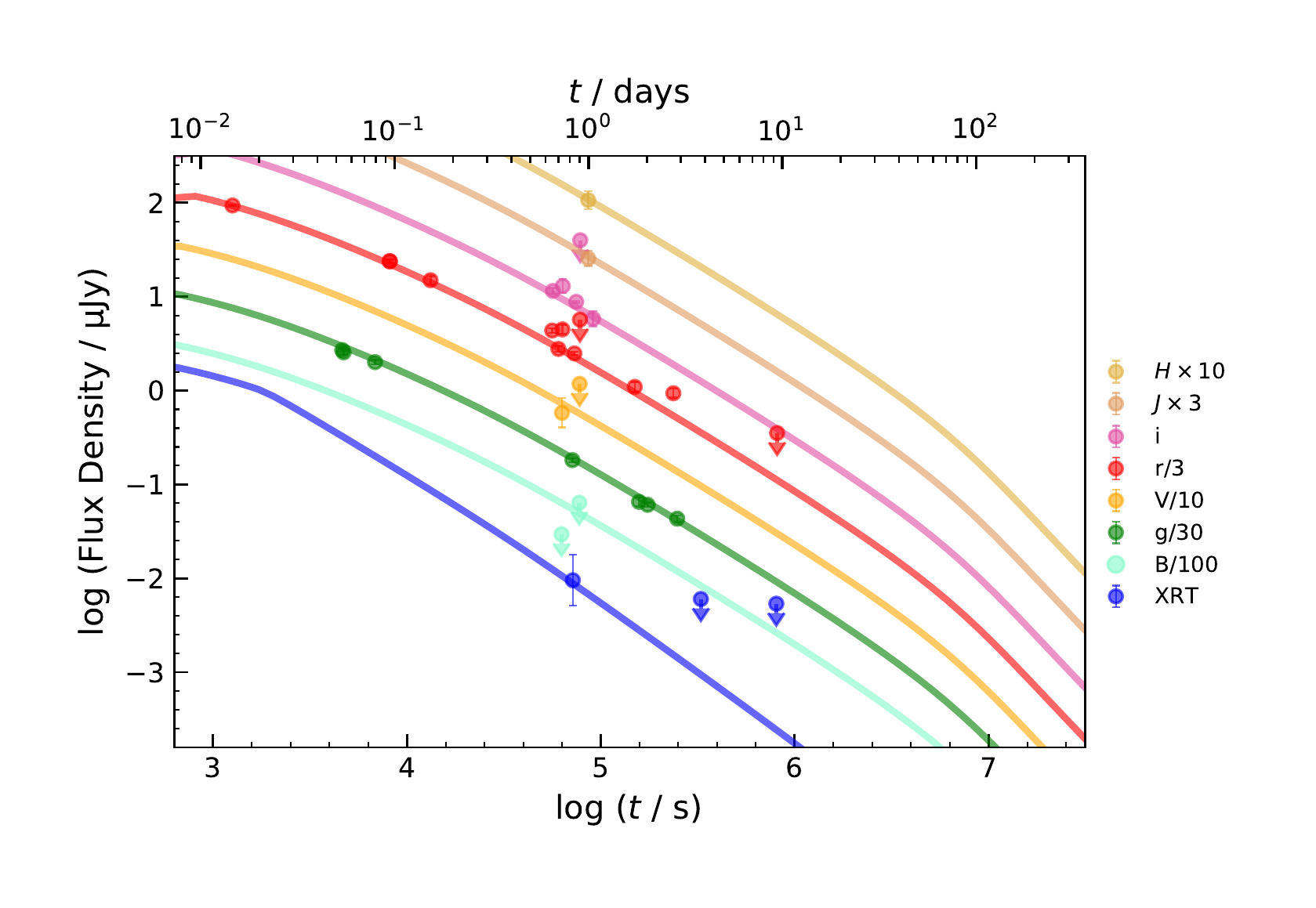}
	}
	
	\caption{Observed and theoretical optical and X-ray afterglow of AT2021any.
		The solid curves represent the best-fit results obtained from optical and X-ray data by using the top-hat jet model.} \label{fig:5}
\end{figure*}

\begin{figure*}
	\centering
	\subfloat{
		\includegraphics[width=\textwidth]{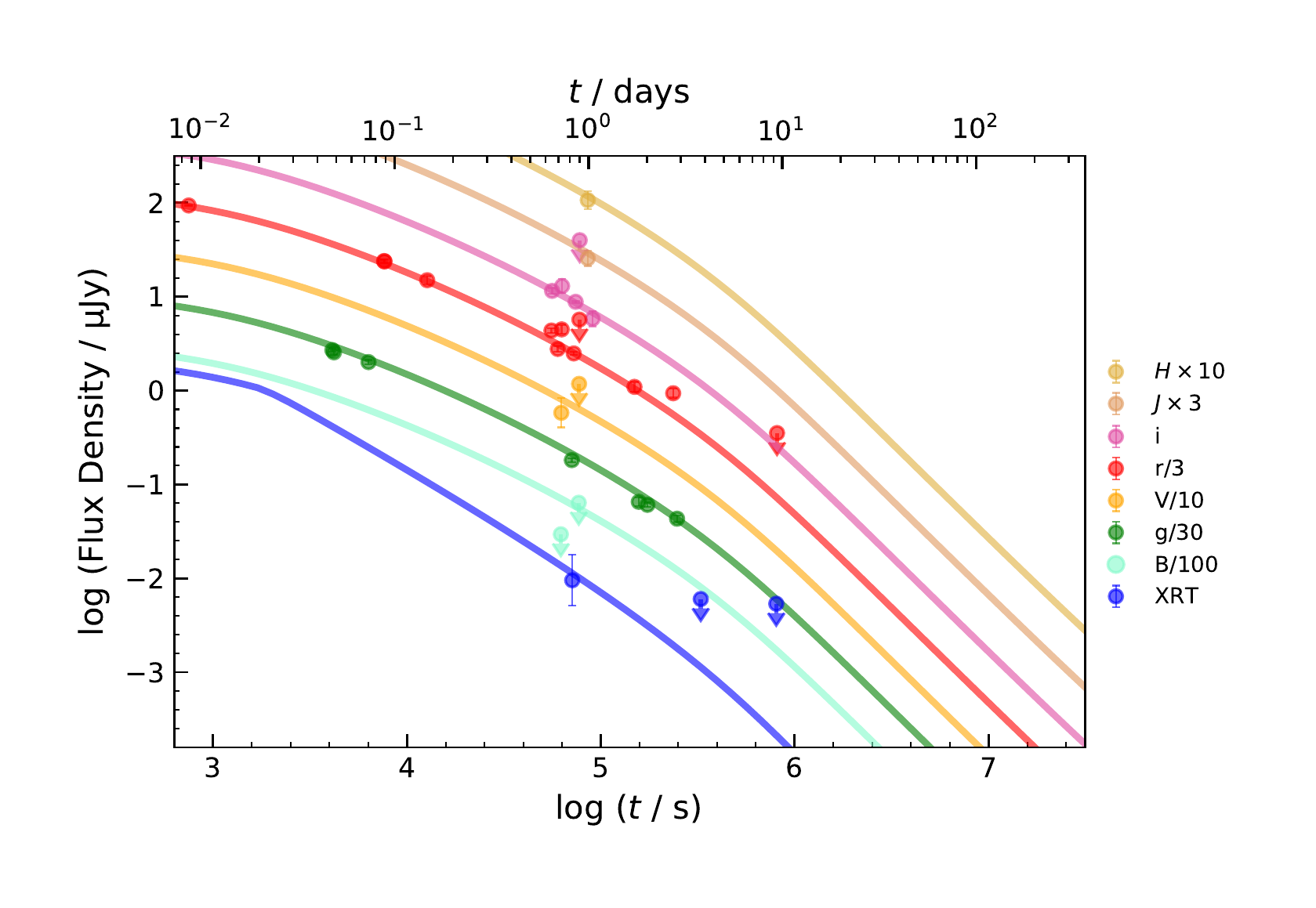}
	}
	
	\caption{Observed and theoretical optical and X-ray afterglow of AT2021any.
		The solid curves represent the best-fit results obtained from optical and X-ray data by using the structured jet model.} \label{fig:6}
\end{figure*}

Radio afterglows are largely affected by interstellar scintillation.
The random fluctuation of radio flux may affect the goodness of fitting.
To avoid the uncertainties caused by this factor, we have performed the
model fitting by excluding all the radio data. The parameters derived are
also presented in Table \ref{tab:4}. Figure \ref{fig:5} and Figure \ref{fig:6}
show the best-fit light curves for the top-hat and structured jet models,
respectively. For the top-hat jet model, the geometry parameters differ
significantly from the previous results when the radio data are included.
The best-fit angles are $\jet = 0.17^{+0.03}_{-0.04}$
and $\obs = 0.12^{+0.04}_{-0.04}$ now, which are relatively larger. Still,
an on-axis scenario is favored. Other parameters such
as $\Gamma_{0}, \Ek, p, n, \epsilon_{\rm e}, \epsilon_{\rm B},
t_{\rm s}$ do not change too much. As for the structured jet model, the major
difference induced by excluding the radio data is about the shift time. We get
the new shift time as $\ts \sim 750$ s, which is $250$ s smaller than the previous
value derived by including the radio data. Apart from this difference, the
best-fit values for the other parameters are essentially similar. Finally,
from both the reduce-$\chi^{2}$ test and the BIC test, the structured jet model
is still preferred after excluding the radio data, thus the main conclusion
remains unchanged. 

To summarize, according to the above two tests, AT2021any is best described
by the structured jet model. This conclusion is supported either the radio
data, which are affected by interstellar scintillation, are included or
excluded. Additionally, our results suggest that AT2021any should be
an on-axis failed GRB.

\section{Conclusions and discussion} \label{sec:dis}	

Searching for orphan afterglows is a difficult but meaningful task. By fitting
the multi-wavelength afterglow data, we will get a better understanding of
the physics of orphan afterglows. In this study, two different kinds of
outflows are applied to the orphan afterglow candidate of AT2021any, i.e. a
top-hat jet and a structured Gaussian jet. It is
found that the structured Gaussian jet model presents the best fit to the
multi-wavelength light curves of AT2021any. According to our modeling, the
trigger time of the GRB associated with AT2021any is about $1000$ s prior
to the first detection.

In the framework of the structured Gaussian jet model, the isotropic
kinetic energy is derived as $5.50 \times 10^{52}$ erg. From this kinetic
energy, we could estimate the $\gamma$-ray efficiency $\eta$ of the unseen
GRB associated with AT2021any. The source was in the field of view of
Fermi-GBM but was undetected by the instrument \citep{Ho..2022}, which places
a firm upper limit on the peak $\gamma$-ray flux
as $\sim 1 \times 10^{7}$ erg s$^{-1}$ cm$^{-2}$. Then the corresponding
upper limit of the peak luminosity is $L_{\rm{p}} \lesssim 5.31 \times
10^{51}$ erg s$^{-1}$ for a redshift of $z=2.513$. The upper limit of the
isotropic $\gamma$-ray energy will be $E_{\gamma, \rm{iso}} =
L_{\rm{p}}*T_{90}/(1+z) \lesssim 1.51 \times 10^{52}$ erg for a typical
burst duration of $T_{90}=10$ s (for long GRBs). So, we can get the
$\gamma$-ray efficiency as $\eta = E_{\gamma, \rm{iso}} / (E_{\gamma, \rm{iso}} + \Ek)
\lesssim 21.5 \%$, which is roughly consistent with the result derived
by \cite{Gupta..2022} ($\eta \lesssim 28.6 \%$). Note that the $\gamma$-ray
efficiency derived here is typical for long GRBs. Some long GRBs
could even have much higher $\gamma$-ray efficiencies but still could be
explained by the photosphere model \citep{Rees..2005,Peer..2008} or the
internal-collision-induced magnetic reconnection and turbulence (ICMART) model \citep{Zhang..2011,Zhang..2014}.

A relativistic fireball with a lower initial Lorentz factor is usually
optically thick at the internal shock radius, but it becomes optical thin at
the external radius. In other words, the photosphere radius of a dirty fireball
is much larger than the internal shock radius and is much smaller than the
external shock radius. In the case of AT2021any, we
have $\Gamma_{0} \sim 68$, $\Ek \sim 5.50 \times 10^{52}$ erg,
and $n \sim 0.87$ cm$^{-3}$ as derived from the structured Gaussian jet model.
Let us consider two mini-shells ejected with a separation time
of $\delta t \sim 10^{-3}$ s. The internal shock radius can be obtained
as $R_{\rm{IS}} \sim 2\Gamma_{0}^{2}c\delta t = 2.69 \times 10^{11}$
cm \citep{Rees..1994}. The optical depth is dominated by electron scattering
for $\Gamma_{0} < 10^{5}$ \citep{Meszaros..1993}, which can be calculated
as $\tau_{\rm{T}} = \frac{\Ek \sigma_{\rm{T}}}{8\pi N_{\rm{sh}}
R \delta R c^{2} m_{\rm{p}} \Gamma_{0}^{3}} $ \citep{Meszaros..2000,Zhang..2018}.
Here $\sigma_{\rm{T}}$ is the electron Thomson cross-section,
$N_{\rm{sh}}$ stands for the number of all the mini-shells,
and $\delta R \sim 3 \times 10^{7}$ cm is their typical width. Taking the
burst duration as $T_{90} = 10$ s and
assuming $N_{\rm{sh}} \sim T_{90}/\delta t = 10^{4}$, we can get the
photosphere radius as $R_{\rm{ph}} = \frac{\Ek \sigma_{\rm{T}}}{8\pi N_{\rm{sh}} \delta R c^{2} m_{\rm{p}} \Gamma_{0}^{3}} = 1.09 \times 10^{13}$ cm for $\tau_{\rm{T}} = 1$. At the same time, the external shock radius
is $R_{\rm{ext}} = \left(\frac{ 3 \Ek }{2\pi n m_{\rm{p}} c^{2} \Gamma_{0}^{2}} \right)^{1/3} = 1.65 \times 10^{17}$ cm \citep{Rees..1992,Zhang..2018}. Comparing these radii, we find that they
satisfy $R_{\rm{IS}} < R_{\rm{ph}} < R_{\rm{ext}}$. It indicates that for
AT2021any, the synchrotron radiation of $\gamma$-rays in the prompt emission 
phase is invisible to us due to the optically thick condition, which is 
consistent with observational constraints. However, emission in the afterglow
phase could be observed since it is optically thin at late stages.

Our model does not include the impact of a reverse shock. Usually, an
optical orphan afterglow would be found at the early stage of the burst,
when the smoking gun is still relatively bright. At this stage, the optical
emission of the afterglow may be affected by the reverse
shock \citep{Wu..2003,Wang..2013}. Note that the reverse shock component
may help us better constrain the physical parameters of an orphan afterglow.
\cite{Wang..2013} engaged the reverse shock emission from a post-merger
millisecond magnetar to explain the light curves of PTF11agg, another
orphan afterglow candidate. They argued that the multi-wavelength light
curves can be better fitted by adding a reverse shock component. Anyway,
in the case of AT2021any studied here, no clear evidence supporting the
existence of a reverse shock is spotted. The reason may be that the first
detection is about 1000 s after the trigger, as derived from our best-fit
result. However, the reverse shock is expected to take effect tens of
seconds after the burst.

The lack of information on the host galaxy extinction is another factor
that may affect the goodness of the multi-wavelength fitting.
\cite{Sarin..2022} added two additional parameters in their fitting of
the orphan afterglow candidate AT2020blt to account for the host galaxy
extinction. These two parameters are both related to the hydrogen column
density of the host galaxy \citep{Guver..2009}. On the other hand,
overestimating the host galaxy extinction could distort the light curve.
So, the extinction of the host galaxy is a tricky problem in the
multi-wavelength fitting. It needs to be considered cautiously.

\begin{acknowledgements}
We would like to thank the anonymous referee for helpful suggestions that lead to an overall improvement of this study.
This study is supported
by the National Natural Science Foundation of China (Grant Nos. 12233002, 12041306, 12147103, U1938201, 12273113),
by National SKA Program of China No. 2020SKA0120300,
by National Key R\&D Program of China (2021YFA0718500),
by the Youth Innovations and Talents Project of Shandong Provincial Colleges and Universities (Grant No. 201909118),
and by the Youth Innovation Promotion Association (2023331).
\end{acknowledgements}

\nocite{*}
\bibliographystyle{aa}
\bibliography{bibtex}

\begin{thebibliography}{87}
\expandafter\ifx\csname natexlab\endcsname\relax\def\natexlab#1{#1}\fi

\bibitem[{{Alexander} {et~al.}(2017){Alexander}, {Berger}, {Fong}, {Williams},
  {Guidorzi}, {Margutti}, {Metzger}, {Annis}, {Blanchard}, {Brout}, {Brown},
  {Chen}, {Chornock}, {Cowperthwaite}, {Drout}, {Eftekhari}, {Frieman}, {Holz},
  {Nicholl}, {Rest}, {Sako}, {Soares-Santos}, \& {Villar}}]{Alexander..2017}
{Alexander}, K.~D., {Berger}, E., {Fong}, W., {et~al.} 2017, \apjl, 848, L21

\bibitem[{{An} {et~al.}(2023){An}, {Antier}, {Bi}, {Bu}, {Cai}, {Cao},
  {Camisasca}, {Chang}, {Chen}, {Chen}, {Chen}, {Chen}, {Chen}, {Chen}, {Chen},
  {Coughlin}, {Cui}, {Dai}, {Hussenot-Desenonges}, {Du}, {Du}, {Du}, {Fan},
  {Frontera}, {Gao}, {Gao}, {Ge}, {Gong}, {Gu}, {Guan}, {Guo}, {Guo},
  {Guidorzi}, {Han}, {He}, {He}, {Hou}, {Huang}, {Huo}, {Ji}, {Jia}, {Jiang},
  {Kann}, {Klotz}, {Kong}, {Lan}, {Li}, {Li}, {Li}, {Li}, {Li}, {Li}, {Li},
  {Li}, {Li}, {Li}, {Li}, {Li}, {Li}, {Liang}, {Liang}, {Liao}, {Lin}, {Liu},
  {Liu}, {Liu}, {Liu}, {Liu}, {Liu}, {Liu}, {Lu}, {Lu}, {Lu}, {Luo}, {Luo},
  {Ma}, {Ma}, {Ma}, {Ma}, {Maccary}, {Mao}, {Meng}, {Nie}, {Orlandini}, {Ou},
  {Peng}, {Peng}, {Qiao}, {Qu}, {Ren}, {Shi}, {Shi}, {Song}, {Song}, {Su},
  {Sun}, {Sun}, {Sun}, {Tan}, {Tan}, {Tao}, {Tuo}, {Turpin}, {Wang}, {Wang},
  {Wang}, {Wang}, {Wang}, {Wang}, {Wang}, {Wang}, {Wang}, {Wang}, {Wang},
  {Wang}, {Wang}, {Wang}, {Wen}, {Wu}, {Wu}, {Wu}, {Xiao}, {Xiao}, {Xiao},
  {Xie}, {Xiong}, {Xiong}, {Xu}, {Xu}, {Xu}, {Xu}, {Xu}, {Xu}, {Xue}, {Yang},
  {Yang}, {Yang}, {Ye}, {Yi}, {Yi}, {Yin}, {You}, {Yu}, {Yu}, {Yu}, {Zeng},
  {Zhang}, {Zhang}, {Zhang}, {Zhang}, {Zhang}, {Zhang}, {Zhang}, {Zhang},
  {Zhang}, {Zhang}, {Zhang}, {Zhang}, {Zhang}, {Zhang}, {Zhang}, {Zhang},
  {Zhang}, {Zhang}, {Zhang}, {Zhao}, {Zhao}, {Zhao}, {Zhao}, {Zhao}, {Zhao},
  {Zhao}, {Zhao}, {Zheng}, {Zheng}, {Zhou}, {Zhou}, \& {Zhu}}]{An..2023}
{An}, Z.-H., {Antier}, S., {Bi}, X.-Z., {et~al.} 2023, arXiv e-prints,
  arXiv:2303.01203

\bibitem[{{Andreoni} {et~al.}(2021){Andreoni}, {Coughlin}, {Kool}, {Kasliwal},
  {Kumar}, {Bhalerao}, {Carracedo}, {Ho}, {Pang}, {Saraogi}, {Sharma},
  {Shenoy}, {Burns}, {Ahumada}, {Anand}, {Singer}, {Perley}, {De}, {Fremling},
  {Bellm}, {Bulla}, {Crellin-Quick}, {Dietrich}, {Drake}, {Duev}, {Goobar},
  {Graham}, {Kaplan}, {Kulkarni}, {Laher}, {Mahabal}, {Shupe}, {Sollerman},
  {Walters}, \& {Yao}}]{Andreoni..2021}
{Andreoni}, I., {Coughlin}, M.~W., {Kool}, E.~C., {et~al.} 2021, \apj, 918, 63

\bibitem[{{Bellm} {et~al.}(2019){Bellm}, {Kulkarni}, {Barlow}, {Feindt},
  {Graham}, {Goobar}, {Kupfer}, {Ngeow}, {Nugent}, {Ofek}, {Prince}, {Riddle},
  {Walters}, \& {Ye}}]{Bellm..2019}
{Bellm}, E.~C., {Kulkarni}, S.~R., {Barlow}, T., {et~al.} 2019, \pasp, 131,
  068003

\bibitem[{{Berger} {et~al.}(2013){Berger}, {Leibler}, {Chornock}, {Rest},
  {Foley}, {Soderberg}, {Price}, {Burgett}, {Chambers}, {Flewelling}, {Huber},
  {Magnier}, {Metcalfe}, {Stubbs}, \& {Tonry}}]{Berger..2013}
{Berger}, E., {Leibler}, C.~N., {Chornock}, R., {et~al.} 2013, \apj, 779, 18

\bibitem[{{Cenko} {et~al.}(2013){Cenko}, {Kulkarni}, {Horesh}, {Corsi}, {Fox},
  {Carpenter}, {Frail}, {Nugent}, {Perley}, {Gruber}, {Gal-Yam}, {Groot},
  {Hallinan}, {Ofek}, {Rau}, {MacLeod}, {Miller}, {Bloom}, {Filippenko},
  {Kasliwal}, {Law}, {Morgan}, {Polishook}, {Poznanski}, {Quimby}, {Sesar},
  {Shen}, {Silverman}, \& {Sternberg}}]{Cenko..2013}
{Cenko}, S.~B., {Kulkarni}, S.~R., {Horesh}, A., {et~al.} 2013, \apj, 769, 130

\bibitem[{{Chevalier} \& {Li}(1999)}]{Chevalier..1999}
{Chevalier}, R.~A. \& {Li}, Z.-Y. 1999, \apjl, 520, L29

\bibitem[{{Cordes} \& {Lazio}(2002)}]{Cordes..2002}
{Cordes}, J.~M. \& {Lazio}, T.~J.~W. 2002, arXiv e-prints, astro

\bibitem[{{Dai} \& {Lu}(2001)}]{Dai..2001}
{Dai}, Z.~G. \& {Lu}, T. 2001, \apj, 551, 249

\bibitem[{{de Ugarte Postigo} {et~al.}(2021){de Ugarte Postigo}, {Kann},
  {Perley}, {Thoene}, {Blazek}, {Agui Fernandez}, \& {Scarpa}}]{Postigo..2021}
{de Ugarte Postigo}, A., {Kann}, D.~A., {Perley}, D.~A., {et~al.} 2021, GRB
  Coordinates Network, 29307, 1

\bibitem[{{Dermer} {et~al.}(1999){Dermer}, {Chiang}, \&
  {B{\"o}ttcher}}]{Dermer..1999}
{Dermer}, C.~D., {Chiang}, J., \& {B{\"o}ttcher}, M. 1999, \apj, 513, 656

\bibitem[{{Fishman} \& {Meegan}(1995)}]{Fishman..1995}
{Fishman}, G.~J. \& {Meegan}, C.~A. 1995, \araa, 33, 415

\bibitem[{{Gal-Yam} {et~al.}(2006){Gal-Yam}, {Ofek}, {Poznanski}, {Levinson},
  {Waxman}, {Frail}, {Soderberg}, {Nakar}, {Li}, \& {Filippenko}}]{Gal..2006}
{Gal-Yam}, A., {Ofek}, E.~O., {Poznanski}, D., {et~al.} 2006, \apj, 639, 331

\bibitem[{{Gao} {et~al.}(2013){Gao}, {Lei}, {Zou}, {Wu}, \&
  {Zhang}}]{Gao..2013}
{Gao}, H., {Lei}, W.-H., {Zou}, Y.-C., {Wu}, X.-F., \& {Zhang}, B. 2013, \nar,
  57, 141

\bibitem[{{Gao} {et~al.}(2022){Gao}, {Geng}, {Hu}, {Hu}, {Lan}, {Chang},
  {Zhang}, {Zhang}, {Huang}, \& {Wu}}]{Gao..2022}
{Gao}, H.-X., {Geng}, J.-J., {Hu}, L., {et~al.} 2022, \mnras, 516, 453

\bibitem[{{Geng} {et~al.}(2016){Geng}, {Wu}, {Huang}, {Li}, \&
  {Dai}}]{Geng..2016}
{Geng}, J.~J., {Wu}, X.~F., {Huang}, Y.~F., {Li}, L., \& {Dai}, Z.~G. 2016,
  \apj, 825, 107

\bibitem[{{Geng} {et~al.}(2013){Geng}, {Wu}, {Huang}, \& {Yu}}]{Geng..2013}
{Geng}, J.~J., {Wu}, X.~F., {Huang}, Y.~F., \& {Yu}, Y.~B. 2013, \apj, 779, 28

\bibitem[{{Geng} {et~al.}(2019){Geng}, {Zhang}, {K{\"o}lligan}, {Kuiper}, \&
  {Huang}}]{Geng..2019}
{Geng}, J.-J., {Zhang}, B., {K{\"o}lligan}, A., {Kuiper}, R., \& {Huang}, Y.-F.
  2019, \apjl, 877, L40

\bibitem[{{Greiner} {et~al.}(2000){Greiner}, {Hartmann}, {Voges}, {Boller},
  {Schwarz}, \& {Zharikov}}]{Greiner..2000}
{Greiner}, J., {Hartmann}, D.~H., {Voges}, W., {et~al.} 2000, \aap, 353, 998

\bibitem[{{Grindlay}(1999)}]{Grindlay..1999}
{Grindlay}, J.~E. 1999, \apj, 510, 710

\bibitem[{{Gruber} {et~al.}(2014){Gruber}, {Goldstein}, {Weller von Ahlefeld},
  {Narayana Bhat}, {Bissaldi}, {Briggs}, {Byrne}, {Cleveland}, {Connaughton},
  {Diehl}, {Fishman}, {Fitzpatrick}, {Foley}, {Gibby}, {Giles}, {Greiner},
  {Guiriec}, {van der Horst}, {von Kienlin}, {Kouveliotou}, {Layden}, {Lin},
  {Meegan}, {McGlynn}, {Paciesas}, {Pelassa}, {Preece}, {Rau}, {Wilson-Hodge},
  {Xiong}, {Younes}, \& {Yu}}]{Gruber..2014}
{Gruber}, D., {Goldstein}, A., {Weller von Ahlefeld}, V., {et~al.} 2014, \apjs,
  211, 12

\bibitem[{{Gupta} {et~al.}(2022){Gupta}, {Kumar}, {Pandey}, {Castro-Tirado},
  {Ghosh}, {Dimple}, {Fern{\'a}ndez-Garc{\'\i}a}, {Caballero-Garc{\'\i}a},
  {Castro-Tirado}, {Hedrosa}, {Hermelo}, {Vico}, {Misra}, {Kumar}, {Aryan}, \&
  {Tiwari}}]{Gupta..2022}
{Gupta}, R., {Kumar}, A., {Pandey}, S.~B., {et~al.} 2022, Journal of
  Astrophysics and Astronomy, 43, 11

\bibitem[{{G{\"u}ver} \& {{\"O}zel}(2009)}]{Guver..2009}
{G{\"u}ver}, T. \& {{\"O}zel}, F. 2009, \mnras, 400, 2050

\bibitem[{{Ho} {et~al.}(2018){Ho}, {Kulkarni}, {Nugent}, {Zhao}, {Rusu},
  {Cenko}, {Ravi}, {Kasliwal}, {Perley}, {Adams}, {Bellm}, {Brady}, {Fremling},
  {Gal-Yam}, {Kann}, {Kaplan}, {Laher}, {Masci}, {Ofek}, {Sollerman}, \&
  {Urban}}]{Ho..2018}
{Ho}, A. Y.~Q., {Kulkarni}, S.~R., {Nugent}, P.~E., {et~al.} 2018, \apjl, 854,
  L13

\bibitem[{{Ho} {et~al.}(2020){Ho}, {Perley}, {Beniamini}, {Cenko}, {Kulkarni},
  {Andreoni}, {Singer}, {De}, {Kasliwal}, {Fremling}, {Bellm}, {Dekany},
  {Delacroix}, {Duev}, {Goldstein}, {Golkhou}, {Goobar}, {Graham}, {Hale},
  {Kupfer}, {Laher}, {Masci}, {Miller}, {Neill}, {Riddle}, {Rusholme}, {Shupe},
  {Smith}, {Sollerman}, \& {van Roestel}}]{Ho..2020}
{Ho}, A. Y.~Q., {Perley}, D.~A., {Beniamini}, P., {et~al.} 2020, \apj, 905, 98

\bibitem[{{Ho} {et~al.}(2021){Ho}, {Perley}, {Yao}, {Andreoni}, \& {Zwicky
  Transient Facility Collaboration}}]{Ho..2021a}
{Ho}, A.~Y.~Q., {Perley}, D.~A., {Yao}, Y., {Andreoni}, I., \& {Zwicky
  Transient Facility Collaboration}. 2021, GRB Coordinates Network, 29305, 1

\bibitem[{{Ho} {et~al.}(2022){Ho}, {Perley}, {Yao}, {Svinkin}, {de Ugarte
  Postigo}, {Perley}, {Kann}, {Burns}, {Andreoni}, {Bellm}, {Bissaldi},
  {Bloom}, {Brink}, {Dekany}, {Drake}, {Ag{\"u}{\'\i} Fern{\'a}ndez},
  {Filippenko}, {Frederiks}, {Graham}, {Hristov}, {Kasliwal}, {Kulkarni},
  {Kumar}, {Laher}, {Lysenko}, {Mailyan}, {Malacaria}, {Miller}, {Poolakkil},
  {Riddle}, {Ridnaia}, {Rusholme}, {Savchenko}, {Sollerman}, {Th{\"o}ne},
  {Tsvetkova}, {Ulanov}, \& {von Kienlin}}]{Ho..2022}
{Ho}, A. Y.~Q., {Perley}, D.~A., {Yao}, Y., {et~al.} 2022, \apj, 938, 85

\bibitem[{{Ho} \& {Zwicky Transient Facility Collaboration}(2021)}]{Ho..2021b}
{Ho}, A.~Y.~Q. \& {Zwicky Transient Facility Collaboration}. 2021, GRB
  Coordinates Network, 29313, 1

\bibitem[{{Huang} {et~al.}(2006){Huang}, {Cheng}, \& {Gao}}]{Huang..2006}
{Huang}, Y.~F., {Cheng}, K.~S., \& {Gao}, T.~T. 2006, \apj, 637, 873

\bibitem[{{Huang} {et~al.}(2000{\natexlab{a}}){Huang}, {Dai}, \&
  {Lu}}]{Huang..2000a}
{Huang}, Y.~F., {Dai}, Z.~G., \& {Lu}, T. 2000{\natexlab{a}}, \mnras, 316, 943

\bibitem[{{Huang} {et~al.}(2002){Huang}, {Dai}, \& {Lu}}]{Huang..2002}
{Huang}, Y.~F., {Dai}, Z.~G., \& {Lu}, T. 2002, \mnras, 332, 735

\bibitem[{{Huang} {et~al.}(2000{\natexlab{b}}){Huang}, {Gou}, {Dai}, \&
  {Lu}}]{Huang..2000b}
{Huang}, Y.~F., {Gou}, L.~J., {Dai}, Z.~G., \& {Lu}, T. 2000{\natexlab{b}},
  \apj, 543, 90

\bibitem[{{Huang} {et~al.}(2020){Huang}, {Urata}, {Huang}, {Lee}, {Tsai},
  {Shirasaki}, {Sawicki}, {Arnouts}, {Moutard}, {Gwyn}, {Wang}, {Foucaud},
  {Asada}, {Huber}, {Wainscoat}, \& {Chambers}}]{Huang..2020}
{Huang}, Y.-J., {Urata}, Y., {Huang}, K., {et~al.} 2020, \apj, 897, 69

\bibitem[{{Kann} {et~al.}(2010){Kann}, {Klose}, {Zhang}, {Malesani}, {Nakar},
  {Pozanenko}, {Wilson}, {Butler}, {Jakobsson}, {Schulze}, {Andreev},
  {Antonelli}, {Bikmaev}, {Biryukov}, {B{\"o}ttcher}, {Burenin}, {Castro
  Cer{\'o}n}, {Castro-Tirado}, {Chincarini}, {Cobb}, {Covino}, {D'Avanzo},
  {D'Elia}, {Della Valle}, {de Ugarte Postigo}, {Efimov}, {Ferrero}, {Fugazza},
  {Fynbo}, {G{\r{a}}lfalk}, {Grundahl}, {Gorosabel}, {Gupta}, {Guziy},
  {Hafizov}, {Hjorth}, {Holhjem}, {Ibrahimov}, {Im}, {Israel}, {Je{\'l}inek},
  {Jensen}, {Karimov}, {Khamitov}, {Kizilo{\v{g}}lu}, {Klunko}, {Kub{\'a}nek},
  {Kutyrev}, {Laursen}, {Levan}, {Mannucci}, {Martin}, {Mescheryakov},
  {Mirabal}, {Norris}, {Ovaldsen}, {Paraficz}, {Pavlenko}, {Piranomonte},
  {Rossi}, {Rumyantsev}, {Salinas}, {Sergeev}, {Sharapov}, {Sollerman},
  {Stecklum}, {Stella}, {Tagliaferri}, {Tanvir}, {Telting}, {Testa}, {Updike},
  {Volnova}, {Watson}, {Wiersema}, \& {Xu}}]{Kann..2010}
{Kann}, D.~A., {Klose}, S., {Zhang}, B., {et~al.} 2010, \apj, 720, 1513

\bibitem[{{Khabibullin} {et~al.}(2012){Khabibullin}, {Sazonov}, \&
  {Sunyaev}}]{Khabibullin..2012}
{Khabibullin}, I., {Sazonov}, S., \& {Sunyaev}, R. 2012, \mnras, 426, 1819

\bibitem[{{Kool} {et~al.}(2019){Kool}, {Stein}, {Sharma}, {Karambelkar},
  {Kasliwal}, {Perley}, {Brinnel}, {Nordin}, {Anand}, {Coughlin}, {Singer},
  {Andreoni}, {Waratkar}, {Kumar}, {Khandagale}, {Deshmukh}, {Bhalerao},
  {Anupama}, {Dobie}, {Cenko}, {Ahmuda}, {Bellm}, {Kong}, {Franckowiak},
  {Gatkine}, {Ztf Collaboration}, \& {Growth Collaboration}}]{Kool..2019}
{Kool}, E., {Stein}, R., {Sharma}, Y., {et~al.} 2019, GRB Coordinates Network,
  25616, 1

\bibitem[{{Kumar} \& {Granot}(2003)}]{Kumar..2003}
{Kumar}, P. \& {Granot}, J. 2003, \apj, 591, 1075

\bibitem[{{Kumar} \& {Zhang}(2015)}]{Kumar..2015}
{Kumar}, P. \& {Zhang}, B. 2015, \physrep, 561, 1

\bibitem[{{Lamb} {et~al.}(2021){Lamb}, {Fern{\'a}ndez}, {Hayes}, {Kong}, {Lin},
  {Tanvir}, {Hendry}, {Heng}, {Saha}, \& {Veitch}}]{Lamb..2021}
{Lamb}, G.~P., {Fern{\'a}ndez}, J.~J., {Hayes}, F., {et~al.} 2021, Universe, 7,
  329

\bibitem[{{Law} {et~al.}(2018){Law}, {Gaensler}, {Metzger}, {Ofek}, \&
  {Sironi}}]{Law..2018}
{Law}, C.~J., {Gaensler}, B.~M., {Metzger}, B.~D., {Ofek}, E.~O., \& {Sironi},
  L. 2018, \apjl, 866, L22

\bibitem[{{Law} {et~al.}(2009){Law}, {Kulkarni}, {Dekany}, {Ofek}, {Quimby},
  {Nugent}, {Surace}, {Grillmair}, {Bloom}, {Kasliwal}, {Bildsten}, {Brown},
  {Cenko}, {Ciardi}, {Croner}, {Djorgovski}, {van Eyken}, {Filippenko}, {Fox},
  {Gal-Yam}, {Hale}, {Hamam}, {Helou}, {Henning}, {Howell}, {Jacobsen},
  {Laher}, {Mattingly}, {McKenna}, {Pickles}, {Poznanski}, {Rahmer}, {Rau},
  {Rosing}, {Shara}, {Smith}, {Starr}, {Sullivan}, {Velur}, {Walters}, \&
  {Zolkower}}]{Law..2009}
{Law}, N.~M., {Kulkarni}, S.~R., {Dekany}, R.~G., {et~al.} 2009, \pasp, 121,
  1395

\bibitem[{{Leung} {et~al.}(2023){Leung}, {Murphy}, {Lenc}, {Edwards},
  {Ghirlanda}, {Kaplan}, {O'Brien}, \& {Wang}}]{Leung..2023}
{Leung}, J.~K., {Murphy}, T., {Lenc}, E., {et~al.} 2023, \mnras, 523, 4029

\bibitem[{{Levinson} {et~al.}(2002){Levinson}, {Ofek}, {Waxman}, \&
  {Gal-Yam}}]{Levinson..2002}
{Levinson}, A., {Ofek}, E.~O., {Waxman}, E., \& {Gal-Yam}, A. 2002, \apj, 576,
  923

\bibitem[{{Lipunov} {et~al.}(2022){Lipunov}, {Kornilov}, {Zhirkov}, {Tyurina},
  {Gorbovskoy}, {Vlasenko}, {Simakov}, {Topolev}, {Francile}, {Podesta},
  {Podesta}, {Svinkin}, {Budnev}, {Gress}, {Balanutsa}, {Kuznetsov},
  {Chasovnikov}, {Serra-Ricart}, {Gabovich}, {Minkina}, {Antipov}, {Svertilov},
  {Tlatov}, {Senik}, {Tselik}, {Kechin}, \& {Yurkov}}]{Lipunov..2022}
{Lipunov}, V., {Kornilov}, V., {Zhirkov}, K., {et~al.} 2022, \mnras, 516, 4980

\bibitem[{{M{\'e}sz{\'a}ros}(2006)}]{Meszaros..2006}
{M{\'e}sz{\'a}ros}, P. 2006, Reports on Progress in Physics, 69, 2259

\bibitem[{{M{\'e}sz{\'a}ros} {et~al.}(1993){M{\'e}sz{\'a}ros}, {Laguna}, \&
  {Rees}}]{Meszaros..1993}
{M{\'e}sz{\'a}ros}, P., {Laguna}, P., \& {Rees}, M.~J. 1993, \apj, 415, 181

\bibitem[{{M{\'e}sz{\'a}ros} \& {Rees}(1997)}]{Meszaros..1997}
{M{\'e}sz{\'a}ros}, P. \& {Rees}, M.~J. 1997, \apj, 476, 232

\bibitem[{{M{\'e}sz{\'a}ros} \& {Rees}(2000)}]{Meszaros..2000}
{M{\'e}sz{\'a}ros}, P. \& {Rees}, M.~J. 2000, \apj, 530, 292

\bibitem[{{Mooley} {et~al.}(2022){Mooley}, {Margalit}, {Law}, {Perley},
  {Deller}, {Lazio}, {Bietenholz}, {Shimwell}, {Intema}, {Gaensler}, {Metzger},
  {Dong}, {Hallinan}, {Ofek}, \& {Sironi}}]{Mooley..2022}
{Mooley}, K.~P., {Margalit}, B., {Law}, C.~J., {et~al.} 2022, \apj, 924, 16

\bibitem[{{Nakar} {et~al.}(2002){Nakar}, {Piran}, \& {Granot}}]{Nakar..2002}
{Nakar}, E., {Piran}, T., \& {Granot}, J. 2002, \apj, 579, 699

\bibitem[{{O'Connor} {et~al.}(2023){O'Connor}, {Troja}, {Ryan}, {Beniamini},
  {van Eerten}, {Granot}, {Dichiara}, {Ricci}, {Lipunov}, {Gillanders}, {Gill},
  {Moss}, {Anand}, {Andreoni}, {Becerra}, {Buckley}, {Butler}, {Cenko},
  {Chasovnikov}, {Durbak}, {Francile}, {Hammerstein}, {van der Horst},
  {Kasliwal}, {Kouveliotou}, {Kutyrev}, {Lee}, {Srinivasaragavan}, {Topolev},
  {Watson}, {Yang}, \& {Zhirkov}}]{Connor..2023}
{O'Connor}, B., {Troja}, E., {Ryan}, G., {et~al.} 2023, arXiv e-prints,
  arXiv:2302.07906

\bibitem[{{Paczy{\'n}ski}(1998)}]{Paczynski..1998}
{Paczy{\'n}ski}, B. 1998, \apjl, 494, L45

\bibitem[{{Paczy{\'n}ski}(2000)}]{Paczynski..2000}
{Paczy{\'n}ski}, B. 2000, \pasp, 112, 1281

\bibitem[{{Pe'er}(2008)}]{Peer..2008}
{Pe'er}, A. 2008, \apj, 682, 463

\bibitem[{{Perley} {et~al.}(2021){Perley}, {Ho}, {Yao}, \&
  {Perley}}]{Perley..2021}
{Perley}, D.~A., {Ho}, A. Y.~Q., {Yao}, Y., \& {Perley}, R.~A. 2021, GRB
  Coordinates Network, 29343, 1

\bibitem[{{Piran}(2004)}]{Piran..2004}
{Piran}, T. 2004, Reviews of Modern Physics, 76, 1143

\bibitem[{{Rau} {et~al.}(2006){Rau}, {Greiner}, \& {Schwarz}}]{Rau..2006}
{Rau}, A., {Greiner}, J., \& {Schwarz}, R. 2006, \aap, 449, 79

\bibitem[{{Rees} \& {M{\'e}sz{\'a}ros}(1992)}]{Rees..1992}
{Rees}, M.~J. \& {M{\'e}sz{\'a}ros}, P. 1992, \mnras, 258, 41

\bibitem[{{Rees} \& {M{\'e}sz{\'a}ros}(1994)}]{Rees..1994}
{Rees}, M.~J. \& {M{\'e}sz{\'a}ros}, P. 1994, \apjl, 430, L93

\bibitem[{{Rees} \& {M{\'e}sz{\'a}ros}(2005)}]{Rees..2005}
{Rees}, M.~J. \& {M{\'e}sz{\'a}ros}, P. 2005, \apj, 628, 847

\bibitem[{{Ren} {et~al.}(2023){Ren}, {Wang}, {Zhang}, \& {Dai}}]{Ren..2023}
{Ren}, J., {Wang}, Y., {Zhang}, L.-L., \& {Dai}, Z.-G. 2023, \apj, 947, 53

\bibitem[{{Rhoads}(1997)}]{Rhoads..1997}
{Rhoads}, J.~E. 1997, \apjl, 487, L1

\bibitem[{{Sari} \& {Esin}(2001)}]{Sari..2001}
{Sari}, R. \& {Esin}, A.~A. 2001, \apj, 548, 787

\bibitem[{{Sari} {et~al.}(1999){Sari}, {Piran}, \& {Halpern}}]{Sari..1999}
{Sari}, R., {Piran}, T., \& {Halpern}, J.~P. 1999, \apjl, 519, L17

\bibitem[{{Sari} {et~al.}(1998){Sari}, {Piran}, \& {Narayan}}]{Sari..1998}
{Sari}, R., {Piran}, T., \& {Narayan}, R. 1998, \apjl, 497, L17

\bibitem[{{Sarin} {et~al.}(2022){Sarin}, {Hamburg}, {Burns}, {Ashton}, {Lasky},
  \& {Lamb}}]{Sarin..2022}
{Sarin}, N., {Hamburg}, R., {Burns}, E., {et~al.} 2022, \mnras, 512, 1391

\bibitem[{{Sato} {et~al.}(2023){Sato}, {Murase}, {Ohira}, \&
  {Yamazaki}}]{Sato..2023}
{Sato}, Y., {Murase}, K., {Ohira}, Y., \& {Yamazaki}, R. 2023, \mnras, 522, L56

\bibitem[{{Schlafly} \& {Finkbeiner}(2011)}]{Schlafly..2011}
{Schlafly}, E.~F. \& {Finkbeiner}, D.~P. 2011, \apj, 737, 103

\bibitem[{{Schwarz}(1978)}]{Schwarz..1978}
{Schwarz}, G. 1978, Annals of Statistics, 6, 461

\bibitem[{{Troja} {et~al.}(2018){Troja}, {Piro}, {Ryan}, {van Eerten}, {Ricci},
  {Wieringa}, {Lotti}, {Sakamoto}, \& {Cenko}}]{Troja..2018}
{Troja}, E., {Piro}, L., {Ryan}, G., {et~al.} 2018, \mnras, 478, L18

\bibitem[{{Urata} {et~al.}(2015){Urata}, {Huang}, {Yamazaki}, \&
  {Sakamoto}}]{Urata..2015}
{Urata}, Y., {Huang}, K., {Yamazaki}, R., \& {Sakamoto}, T. 2015, \apj, 806,
  222

\bibitem[{{Wang} \& {Dai}(2013)}]{Wang..2013}
{Wang}, L.-J. \& {Dai}, Z.-G. 2013, \apjl, 774, L33

\bibitem[{{Wei} {et~al.}(2006){Wei}, {Yan}, \& {Fan}}]{Wei..2006}
{Wei}, D.~M., {Yan}, T., \& {Fan}, Y.~Z. 2006, \apjl, 636, L69

\bibitem[{{Willingale} {et~al.}(2013){Willingale}, {Starling}, {Beardmore},
  {Tanvir}, \& {O'Brien}}]{Willingale..2013}
{Willingale}, R., {Starling}, R.~L.~C., {Beardmore}, A.~P., {Tanvir}, N.~R., \&
  {O'Brien}, P.~T. 2013, \mnras, 431, 394

\bibitem[{{Wu} {et~al.}(2003){Wu}, {Dai}, {Huang}, \& {Lu}}]{Wu..2003}
{Wu}, X.~F., {Dai}, Z.~G., {Huang}, Y.~F., \& {Lu}, T. 2003, \mnras, 342, 1131

\bibitem[{{Wu} {et~al.}(2014){Wu}, {Gao}, {Ding}, {Zhang}, {Dai}, \&
  {Wei}}]{Wu..2014}
{Wu}, X.-F., {Gao}, H., {Ding}, X., {et~al.} 2014, \apjl, 781, L10

\bibitem[{{Xu} {et~al.}(2022){Xu}, {Geng}, {Wang}, {Li}, \& {Huang}}]{Xu..2022}
{Xu}, F., {Geng}, J.-J., {Wang}, X., {Li}, L., \& {Huang}, Y.-F. 2022, \mnras,
  509, 4916

\bibitem[{{Xu} {et~al.}(2023){Xu}, {Huang}, {Geng}, {Wu}, {Li}, \&
  {Zhang}}]{Xu..2023}
{Xu}, F., {Huang}, Y.-F., {Geng}, J.-J., {et~al.} 2023, \aap, 673, A20

\bibitem[{{Xu} {et~al.}(2021){Xu}, {Tang}, {Geng}, {Wang}, {Wang}, {Kuerban},
  \& {Huang}}]{Xu..2021}
{Xu}, F., {Tang}, C.-H., {Geng}, J.-J., {et~al.} 2021, \apj, 920, 135

\bibitem[{{Xu} {et~al.}(2012){Xu}, {Nagataki}, {Huang}, \& {Lee}}]{Xu..2012}
{Xu}, M., {Nagataki}, S., {Huang}, Y.~F., \& {Lee}, S.~H. 2012, \apj, 746, 49

\bibitem[{{Yamazaki} {et~al.}(2002){Yamazaki}, {Ioka}, \&
  {Nakamura}}]{Yamazaki..2002}
{Yamazaki}, R., {Ioka}, K., \& {Nakamura}, T. 2002, \apjl, 571, L31

\bibitem[{{Yang} {et~al.}(2023){Yang}, {Zhao}, {Yan}, {Wang}, {Zhang}, {An},
  {Cai}, {Li}, {Li}, {Liu}, {Liu}, {Ma}, {Meng}, {Peng}, {Qiao}, {Shao},
  {Song}, {Tan}, {Wang}, {Wang}, {Wen}, {Xiao}, {Xue}, {Yang}, {Yin}, {Zhang},
  {Zhang}, {Zhang}, {Zhang}, {Zheng}, {Zheng}, {Xiong}, \&
  {Zhang}}]{Yang..2023}
{Yang}, J., {Zhao}, X.-H., {Yan}, Z., {et~al.} 2023, \apjl, 947, L11

\bibitem[{{Zhang}(2018)}]{Zhang..2018}
{Zhang}, B. 2018, {The Physics of Gamma-Ray Bursts} (Cambridge University
  Press)

\bibitem[{{Zhang} {et~al.}(2006){Zhang}, {Fan}, {Dyks}, {Kobayashi},
  {M{\'e}sz{\'a}ros}, {Burrows}, {Nousek}, \& {Gehrels}}]{Zhang..2006}
{Zhang}, B., {Fan}, Y.~Z., {Dyks}, J., {et~al.} 2006, \apj, 642, 354

\bibitem[{{Zhang} \& {Yan}(2011)}]{Zhang..2011}
{Zhang}, B. \& {Yan}, H. 2011, \apj, 726, 90

\bibitem[{{Zhang} \& {Zhang}(2014)}]{Zhang..2014}
{Zhang}, B. \& {Zhang}, B. 2014, \apj, 782, 92

\bibitem[{{Zou} {et~al.}(2007){Zou}, {Wu}, \& {Dai}}]{Zou..2007}
{Zou}, Y.~C., {Wu}, X.~F., \& {Dai}, Z.~G. 2007, \aap, 461, 115

\end{thebibliography}

\begin{table*}[h!]
	\renewcommand{\thetable}{\arabic{table}}
	\centering
	\caption{Optical Photometry of AT2021any} \label{tab:1}
	
	\begin{tabular}{ccccc}
		\hline
		\hline
		Date & $\Delta t$ \tablefootmark{a} & Telescope \tablefootmark{b} & Filter & Magnitude \tablefootmark{c}  \\
		(MJD) & (s) &  &  & (AB) \\
		\hline
		59230.2916 & 997.84 & P48   & r     & 17.92 $\pm$ 0.02 \\
		59230.3307 & 4376.08 & P48   & g     & 19.35 $\pm$ 0.05 \\
		59230.3316 & 4453.84 & P48   & g     & 19.41 $\pm$ 0.06 \\
		59230.3563 & 6587.92 & P48   & g     & 19.67 $\pm$ 0.06 \\
		59230.3712 & 7875.28 & P48   & r     & 19.4  $\pm$ 0.05 \\
		59230.3717 & 7918.48 & P48   & r     & 19.41 $\pm$ 0.05 \\
		59230.4303 & 12981.5 & P48   & r     & 19.91 $\pm$ 0.11 \\
		59230.8004 & 55922.3 & DOT   & R     & 21.25 $\pm$ 0.06 \\
		59230.8052 & 56337 & DOT   & I     & 21.36 $\pm$ 0.08 \\
		59230.8765 & 62497.4 & CAHA  & B     & <22.97 \\
		59230.8803 & 60233.7 & CAHA  & V     & 22.17 $\pm$ 0.39 \\
		59230.8838 & 62825.7 & CAHA  & R     & 21.22 $\pm$ 0.12 \\
		59230.8873 & 63128.1 & CAHA  & I     & 21.23 $\pm$ 0.19 \\
		59230.9767 & 63430.5 & NOT   & g     & 22.28 $\pm$ 0.05 \\
		59230.9772 & 71154.6 & GTC   & r     & 21.74 $\pm$ 0.08 \\
		59230.9957 & 72796.2 & NOT   & r     & 21.86 $\pm$ 0.04 \\
		59231.0147 & 74437.8 & NOT   & i     & 21.65 $\pm$ 0.03 \\
		59231.047 & 77228.6 & CAHA  & B     & <22.13 \\
		59231.0504 & 77522.3 & CAHA  & V     & <21.4  \\
		59231.0538 & 77816.1 & CAHA  & R     & <20.96 \\
		59231.0572 & 78109.8 & CAHA  & I     & <20.01 \\
		59231.1475 & 85911.8 & LBT   & J     & 21.62 $\pm$ 0.2 \\
		59231.1475 & 85911.8 & LBT   & H     & 21.36 $\pm$ 0.24 \\
		59231.2062 & 90983.4 & LDT & i     & 22.1  $\pm$ 0.2 \\
		59231.9736 & 149433 & MPG   & g     & 23.39 $\pm$ 0.11 \\
		59232.0096 & 157287 & CAHA  & r     & 22.75 $\pm$ 0.13 \\
		59232.1705 & 174299 & MPG   & g     & 23.47 $\pm$ 0.06 \\
		59233.0184 & 236593 & CAHA  & r     & 22.92 $\pm$ 0.12 \\
		59233.0206 & 247748 & MPG   & g     & 23.84 $\pm$ 0.09 \\
		59239.5469 & 811620 & DOT   & r     & <23.98 \\
		
		\hline
	\end{tabular}
	\tablefoot{ \\
		\tablefoottext{a}{The time is given relative to the estimated trigger time as derived from the structured jet model. \\}
		\tablefoottext{b}{Some details of the relevant telescopes: 48 inch Samuel Oschin
                          Telescope at Palomar Observatory (P48), 3.6m Devasthal Optical Telescope (DOT),
                          2.2m Telescope at Centro Astrono{\'m}ico Hispano en Andaluc{\'i}a (CAHA),
                          2.56m Nordic Optical Telescope (NOT), 10.4m Gran Telescopio Canarias (GTC),
                          Large Binocular Telescope (LBT), 4.3m Lowell Discovery Telescope (LDT), and 2.2m MPG. \\}
		\tablefoottext{c}{Optical photometry data are taken from \cite{Ho..2022} and \cite{Gupta..2022}.}
	}
\end{table*}

\begin{table*}[h!]
	\renewcommand{\thetable}{\arabic{table}}
	\centering
	\caption{X-ray Afterglow Data of AT2021any} \label{tab:2}
	
	\begin{tabular}{ccc}
		\hline
		\hline
		Date & $\Delta t$ \tablefootmark{a}  & Flux \tablefootmark{b}  \\
		(MJD) & (s) & ($10^{-13}$ erg s$^{-1}$ cm$^{-2}$) \\
		\hline
  		59231.1 & 71491.6 & 3.3 $\pm$ 0.9 \\
		59234.1 & 328100 & <2.10 \\
		59239.6 & 804164 & <1.90 \\
		\hline
	\end{tabular}
	\tablefoot{ \\
		\tablefoottext{a}{The time is given relative to the estimated trigger time as derived from the structured jet model. \\}
		\tablefoottext{b}{The X-ray flux data are taken from \cite{Ho..2022}.}
	}
\end{table*}

\begin{table*}[h!]
	\renewcommand{\thetable}{\arabic{table}}
	\centering
	\caption{Radio Afterglow Data of AT2021any} \label{tab:3}
	
	\begin{tabular}{cccc}
		\hline
		\hline
		Data & $\Delta t$ \tablefootmark{a} & Band & Flux \tablefootmark{b}  \\
		(MJD) & (s) &  & ($\rm{\mu Jy}$) \\
		\hline
    	59235.2 & 424003.6 & X     & 91 $\pm$ 5 \\
		59235.2 & 424003.6 & X     & 63 $\pm$ 6 \\
		59235.2 & 424003.6 & X     & 116 $\pm$ 8 \\
		59235.2 & 424003.6 & X     & 66 $\pm$ 9 \\
		59235.2 & 424003.6 & X     & 62 $\pm$ 7 \\
		59235.2 & 424003.6 & X     & 99 $\pm$ 8 \\
		59235.2 & 424003.6 & X     & 133 $\pm$ 9 \\
		59237.1 & 592483.6 & Ku    & 21 $\pm$ 4 \\
		59237.1 & 592483.6 & Ku    & 20 $\pm$ 6 \\
		59237.1 & 592483.6 & Ku    & 24 $\pm$ 6 \\
		59237.1 & 592483.6 & Ku    & 25 $\pm$ 9 \\
		59237.1 & 592483.6 & Ku    & 20 $\pm$ 7 \\
		59237.1 & 592483.6 & Ku    & 32 $\pm$ 8 \\
		59237.1 & 592483.6 & X     & 25 $\pm$ 4 \\
		59237.1 & 592483.6 & X     & 33 $\pm$ 6 \\
		59237.1 & 592483.6 & X     & <15 \\
		59237.1 & 592483.6 & X     & 37 $\pm$ 9 \\
		59237.1 & 592483.6 & X     & 35 $\pm$ 8 \\
		59237.1 & 592483.6 & X     & 29 $\pm$ 8 \\
		59240.2 & 855139.6 & C     & <22 \\
		59240.2 & 855139.6 & C     & 34 $\pm$ 7 \\
		59240.2 & 855139.6 & C     & 30 $\pm$ 5 \\
		59240.2 & 855139.6 & Ku    & 31 $\pm$ 6 \\
		59240.2 & 855139.6 & Ku    & 46 $\pm$ 9 \\
		59240.2 & 855139.6 & S     & 67 $\pm$ 14 \\
		59240.2 & 855139.6 & S     & 38 $\pm$ 11 \\
		59240.2 & 855139.6 & S     & <47 \\
		59240.2 & 855139.6 & X     & <21 \\
		59240.2 & 855139.6 & X     & <27 \\
		59244.3 & 1212835.6 & X     & 24 $\pm$ 5 \\
		59244.3 & 1212835.6 & X     & 32 $\pm$ 4 \\
		59244.3 & 1212835.6 & X     & 16 $\pm$ 4 \\
		59244.3 & 1212835.6 & X     & 34 $\pm$ 6 \\
		59244.3 & 1212835.6 & X     & 32 $\pm$ 6 \\
		59244.3 & 1212835.6 & X     & 20 $\pm$ 6 \\
		59244.3 & 1212835.6 & X     & <24 \\
		59251.1 & 1797763.6 & X     & 78 $\pm$ 6 \\
		59251.1 & 1797763.6 & X     & 89 $\pm$ 6 \\
		59251.1 & 1797763.6 & X     & 67 $\pm$ 9 \\
		59251.1 & 1797763.6 & X     & 98 $\pm$ 8 \\
		59251.1 & 1797763.6 & X     & 78 $\pm$ 7 \\
		59251.1 & 1797763.6 & X     & 72 $\pm$ 11 \\
		59251.1 & 1797763.6 & X     & 43 $\pm$ 15 \\
		59258.3 & 2416387.6 & X     & 13 $\pm$ 4 \\
		59273.1 & 3701155.6 & X     & 18 $\pm$ 3 \\
		59306.1 & 6547171.6 & X     & <12 \\
		59306.1 & 6547171.6 & X     & <17 \\
		\hline
	\end{tabular}
	\tablefoot{ \\
		\tablefoottext{a}{The time is given relative to the estimated trigger time as derived from the structured jet model. \\}
		\tablefoottext{b}{Radio afterglow data are taken from \cite{Ho..2022}.}
	}
\end{table*}

\begin{table*}[h!]
	\renewcommand{\thetable}{\arabic{table}}
	\centering
	\tabcolsep=15pt
	\renewcommand\arraystretch{1.5}
	\caption{The Best-Fit Results for the Top-hat Jet Model and the Structured jet Model and Their Corresponding Goodness of Fitting} \label{tab:4}
	
	\begin{tabular}{c|cc|cc}
		\hline
		& \multicolumn{2}{c|}{Fitting with radio data} & \multicolumn{2}{c}{Fitting without radio data} \\
		\hline
		Parameters & Top-hat jet \tablefootmark{a} & Structured jet & Top-hat jet & Structured jet \\
		\hline
		$\log~\Gamma_{0}$ & 1.92 $^{+0.06}_{-0.05}$  & 1.83 $^{+0.04}_{-0.04}$ & 1.95 $^{+0.12}_{-0.08}$ & 1.81 $^{+0.04}_{-0.04}$ \\
		$\log~(\Ek/\rm{erg})$ & 52.90 $^{+0.12}_{-0.12}$  & 52.74 $^{+0.12}_{-0.13}$ & 53.10 $^{+0.12}_{-0.10}$ & 52.99 $^{+0.15}_{-0.13}$ \\
		$\theta_{\rm{c}}$ &  ...  & 0.10 $^{+0.01}_{-0.01}$ &  ... & 0.11 $^{+0.01}_{-0.01}$ \\
		$\theta_{\rm{j}}$ & 0.08 $^{+0.01}_{-0.01}$  & 0.76 $^{+0.50}_{-0.46}$ & 0.17 $^{+0.03}_{-0.04}$ & 0.80 $^{+0.48}_{-0.46}$ \\
		$\theta_{\rm{obs}}$ & 0.03 $^{+0.01}_{-0.01}$  & 0.02 $^{+0.003}_{-0.002}$ & 0.12 $^{+0.04}_{-0.04}$ & 0.02 $^{+0.003}_{-0.003}$ \\
		$p$ & 2.39 $^{+0.02}_{-0.02}$  & 2.28 $^{+0.05}_{-0.05}$ & 2.41 $^{+0.02}_{-0.02}$ & 2.35 $^{+0.05}_{-0.06}$ \\
		$\log~(n/\rm{cm}^{-3})$ & -0.78 $^{+0.19}_{-0.19}$  & -0.06 $^{+0.26}_{-0.24}$ & -0.73 $^{+0.20}_{-0.18}$ & -0.17 $^{+0.19}_{-0.19}$ \\
		$\log~\epsilon_{\rm{e}}$ & -0.94 $^{+0.05}_{-0.05}$  & -0.77 $^{+0.07}_{-0.06}$ & -1.03 $^{+0.07}_{-0.08}$ & -0.88 $^{+0.07}_{-0.07}$ \\
		$\log~\epsilon_{B}$ & -2.76 $^{+0.24}_{-0.24}$  & -3.03 $^{+0.21}_{-0.22}$ & -2.99 $^{+0.16}_{-0.15}$ & -3.23 $^{+0.16}_{-0.17}$ \\
		$t_{\rm{s}}/\rm{ks}$ & 1.21 $^{+0.23}_{-0.23}$  & 1.00 $^{+0.22}_{-0.22}$ & 1.26 $^{+0.49}_{-0.48}$ & 0.75 $^{+0.31}_{-0.36}$ \\
		\hline
		$\chi^{2}/$ d.o.f. & 167.10/26 & 148.23/25 & 113.60/16 & 103.48/15 \\
		$\Delta$BIC & 9.85 & 0 & 7.62 & 0 \\
		\hline
	\end{tabular}
	\tablefoot{ \\
		\tablefoottext{a}{The best-fit parameters are presented with $1\sigma$ uncertainties.}
	}
\end{table*}

\end{document}